\documentclass[prc,aps,superscriptaddress,twocolumn,floatfix,altaffilletter, nofootinbib]{revtex4-1}

\usepackage{graphicx}
\usepackage{epsfig}
\usepackage{amsmath}
\usepackage{amssymb}
\usepackage{xcolor}
\usepackage{multirow}
\usepackage{tabularx}
\usepackage{hyperref}
\hypersetup{
	colorlinks=true,
	linkcolor=blue,
	urlcolor=black,
	citecolor=blue
}
\begin{document}


\title{In-beam $\gamma$-ray and electron spectroscopy of $^{249,251}$Md}


\author{R.~Briselet}
\affiliation{Irfu, CEA, Universit\'e Paris-Saclay, F-91191 Gif-sur-Yvette, France}
\author{Ch.~Theisen}\email[]{christophe.theisen@cea.fr}
\affiliation{Irfu, CEA, Universit\'e Paris-Saclay, F-91191 Gif-sur-Yvette, France}
\author{B.~Sulignano}
\affiliation{Irfu, CEA, Universit\'e Paris-Saclay, F-91191 Gif-sur-Yvette, France}
\author{M.~Airiau}
\affiliation{Irfu, CEA, Universit\'e Paris-Saclay, F-91191 Gif-sur-Yvette, France}
\author{K.~Auranen}
\affiliation{University of Jyvaskyla, Department of Physics, P.O. Box 35, FI-40014 Jyvaskyla, Finland}
\author{D.~M.~Cox}
\altaffiliation{Present address: University of Lund, Box 118, 221 00 Lund, Sweden}
\affiliation{University of Jyvaskyla, Department of Physics, P.O. Box 35, FI-40014 Jyvaskyla, Finland}
\affiliation{University of Liverpool, Department of Physics, Oliver Lodge Laboratory, Liverpool L69 7ZE, United Kingdom}
\author{F.~D\'echery}
\affiliation{Irfu, CEA, Universit\'e Paris-Saclay, F-91191 Gif-sur-Yvette, France}
\affiliation{Institut Pluridisciplinaire Hubert Curien, F-67037 Strasbourg, France}
\author{A.~Drouart}
\affiliation{Irfu, CEA, Universit\'e Paris-Saclay, F-91191 Gif-sur-Yvette, France}
\author{Z.~Favier}
\affiliation{Irfu, CEA, Universit\'e Paris-Saclay, F-91191 Gif-sur-Yvette, France}
\author{B.~Gall}
\affiliation{Institut Pluridisciplinaire Hubert Curien, F-67037 Strasbourg, France}
\author{T.~Goigoux}
\affiliation{Irfu, CEA, Universit\'e Paris-Saclay, F-91191 Gif-sur-Yvette, France}
\author{T.~Grahn}
\affiliation{University of Jyvaskyla, Department of Physics, P.O. Box 35, FI-40014 Jyvaskyla, Finland}
\author{P.~T.~Greenlees}
\affiliation{University of Jyvaskyla, Department of Physics, P.O. Box 35, FI-40014 Jyvaskyla, Finland}
\author{K.~Hauschild}
\affiliation{Universit\'e Paris-Saclay, CNRS/IN2P3, IJCLab, 91405 Orsay, France}
\author{A.~Herzan}
\altaffiliation{Present address: Institute of Physics, Slovak Academy of Sciences, SK-84511 Bratislava, Slovakia}
\affiliation{University of Jyvaskyla, Department of Physics, P.O. Box 35, FI-40014 Jyvaskyla, Finland}
\author{R.-D.~Herzberg}
\affiliation{University of Liverpool, Department of Physics, Oliver Lodge Laboratory, Liverpool L69 7ZE, United Kingdom}
\author{U.~Jakobsson}
\altaffiliation{Present address: Laboratory of Radiochemistry, Department of Chemistry, P.O. Box 55, FI-00014 University of Helsinki, Finland}
\affiliation{University of Jyvaskyla, Department of Physics, P.O. Box 35, FI-40014 Jyvaskyla, Finland}
\author{R.~Julin}
\affiliation{University of Jyvaskyla, Department of Physics, P.O. Box 35, FI-40014 Jyvaskyla, Finland}
\author{S.~Juutinen}
\affiliation{University of Jyvaskyla, Department of Physics, P.O. Box 35, FI-40014 Jyvaskyla, Finland}
\author{J.~Konki}
\altaffiliation{Present address: CERN, CH-1211 Geneva 23, Switzerland}
\affiliation{University of Jyvaskyla, Department of Physics, P.O. Box 35, FI-40014 Jyvaskyla, Finland}
\author{M.~Leino}
\affiliation{University of Jyvaskyla, Department of Physics, P.O. Box 35, FI-40014 Jyvaskyla, Finland}
\author{A.~Lopez-Martens}
\affiliation{Universit\'e Paris-Saclay, CNRS/IN2P3, IJCLab, 91405 Orsay, France}
\author{A.~Mistry}
\altaffiliation{Present address: GSI Helmholtzzentrum f\"ur Schwerionenforschung GmbH, 64291 Darmstadt, Germany}
\affiliation{University of Liverpool, Department of Physics, Oliver Lodge Laboratory, Liverpool L69 7ZE, United Kingdom}
\author{P.~Nieminen}
\altaffiliation{Present address: Fortum Oyj, Power Division, P.O. Box 100, 00048 Fortum, Finland}
\affiliation{University of Jyvaskyla, Department of Physics, P.O. Box 35, FI-40014 Jyvaskyla, Finland}
\author{J.~Pakarinen}
\affiliation{University of Jyvaskyla, Department of Physics, P.O. Box 35, FI-40014 Jyvaskyla, Finland}
\author{P.~Papadakis}
\altaffiliation{Present address: STFC Daresbury Laboratory, Daresbury, Warrington WA4 4AD, United Kingdom}
\affiliation{University of Jyvaskyla, Department of Physics, P.O. Box 35, FI-40014 Jyvaskyla, Finland}
\affiliation{University of Liverpool, Department of Physics, Oliver Lodge Laboratory, Liverpool L69 7ZE, United Kingdom}
\author{P.~Peura}
\altaffiliation{Present address: International Atomic Energy Agency, Vienna, Austria}
\affiliation{University of Jyvaskyla, Department of Physics, P.O. Box 35, FI-40014 Jyvaskyla, Finland}
\author{E.~Rey-Herme}
\affiliation{Irfu, CEA, Universit\'e Paris-Saclay, F-91191 Gif-sur-Yvette, France}
\author{P.~Rahkila}
\affiliation{University of Jyvaskyla, Department of Physics, P.O. Box 35, FI-40014 Jyvaskyla, Finland}
\author{J.~Rubert}
\affiliation{Institut Pluridisciplinaire Hubert Curien, F-67037 Strasbourg, France}
\author{P.~Ruotsalainen}
\affiliation{University of Jyvaskyla, Department of Physics, P.O. Box 35, FI-40014 Jyvaskyla, Finland}
\author{M.~Sandzelius}
\affiliation{University of Jyvaskyla, Department of Physics, P.O. Box 35, FI-40014 Jyvaskyla, Finland}
\author{J.~Sar\'en}
\affiliation{University of Jyvaskyla, Department of Physics, P.O. Box 35, FI-40014 Jyvaskyla, Finland}
\author{C.~Scholey}
\altaffiliation{Present address: The Manufacturing Technology Centre, Pilot Way, Ansty, CV7 9JU,  United Kingdom}
\affiliation{University of Jyvaskyla, Department of Physics, P.O. Box 35, FI-40014 Jyvaskyla, Finland}
\author{J.~Sorri}
\altaffiliation{Present address: Sodankyl\"a Geophysical Observatory, University of Oulu, 90014 Oulu, Finland}
\affiliation{University of Jyvaskyla, Department of Physics, P.O. Box 35, FI-40014 Jyvaskyla, Finland}
\author{S.~Stolze}
\altaffiliation{Present address: Physics Division, Argonne National Laboratory, 9700 South Cass Avenue, Lemont, Illinois 60439, USA}
\affiliation{University of Jyvaskyla, Department of Physics, P.O. Box 35, FI-40014 Jyvaskyla, Finland}
\author{J.~Uusitalo}
\affiliation{University of Jyvaskyla, Department of Physics, P.O. Box 35, FI-40014 Jyvaskyla, Finland}
\author{M.~Vandebrouck}
\affiliation{Irfu, CEA, Universit\'e Paris-Saclay, F-91191 Gif-sur-Yvette, France}
\author{A.~Ward}
\affiliation{University of Liverpool, Department of Physics, Oliver Lodge Laboratory, Liverpool L69 7ZE, United Kingdom}
\author{M.~Zieli\'nska}
\affiliation{Irfu, CEA, Universit\'e Paris-Saclay, F-91191 Gif-sur-Yvette, France}

\author{B.~Bally}
\altaffiliation{Present address: Departamento de F\'isica Te\'orica, Universidad Aut\'onoma de Madrid, E-28049 Madrid, Spain}
\affiliation{Irfu, CEA, Universit\'e Paris-Saclay, F-91191 Gif-sur-Yvette, France}
\author{M.~Bender}
\affiliation{IP2I Lyon, CNRS/IN2P3, Universit\'e Claude Bernard Lyon 1, 
F-69622, Villeurbanne, France}
\author{W.~Ryssens}
\affiliation{Center for Theoretical Physics, Sloane Physics Laboratory, Yale University, New Haven, Connecticut 06520, USA}



\date{\today}

\begin{abstract}
The odd-$Z$ $^{251}$Md nucleus was studied using combined $\gamma$-ray and conversion-electron in-beam spectroscopy.
Besides the previously observed rotational band based on the $[521]1/2^-$ configuration, another rotational structure
has been identified using $\gamma$-$\gamma$ coincidences.
The use of electron spectroscopy allowed the rotational bands to be observed over a larger rotational frequency range. 
Using the transition intensities that depend on the gyromagnetic factor, a $[514]7/2^-$ single-particle configuration has been inferred for this band, 
i.e., the ground-state band.
A physical background that dominates the electron spectrum with an intensity of $\simeq$ 60\% was well reproduced by simulating
a set of unresolved excited bands.
Moreover, a detailed analysis of the intensity profile as a function of the angular momentum provided
a method for deriving the orbital gyromagnetic factor, namely $g_K = 0.69^{+0.19}_{-0.16}$ for the ground-state band.
The odd-$Z$ $^{249}$Md was studied using $\gamma$-ray in-beam spectroscopy.
Evidence for octupole correlations resulting from the
mixing of the $\Delta l = \Delta j = 3$ $[521]3/2^-$ and $[633]7/2^+$ Nilsson orbitals were found in both $^{249,251}$Md.
A surprising similarity of the $^{251}$Md ground-state band transition
energies with those of the excited band of $^{255}$Lr has been discussed in terms of identical bands.
Skyrme-Hartree-Fock-Bogoliubov calculations were performed to investigate the origin of the similarities between these bands.
\end{abstract}


\maketitle


\section{Introduction}

Despite significant and steady advances in the synthesis of the heaviest elements,
reaching the predicted superheavy island of stability is still a distant objective,
because of the ever-decreasing cross sections.
Nevertheless, nuclear spectroscopy, mass measurements and laser spectroscopy of the heaviest nuclei have shown
their effectiveness by providing information on the quantum nature of extreme mass
nuclei~\cite{Block2010, Theisen2015, Asai2015, Ackermann2017, Raeder2018},
without which the nuclei would no longer be bound beyond $Z \simeq 104$.
On the theoretical side,
the island of enhanced stability has been predicted either around the proton number
$Z$=114, 120, or 126 and neutron number $N$ =172 or 184 ~\cite{Cwiok1996, Rutz1997, Buervenich1998ChT, Bender1999, Dobaczewski2015ChT}. 
The validity of these predictions in a region where the models are extrapolated is hence questionable, as is the concept of magic numbers in this region
\cite{Bender2001}.
It is therefore essential to compare predictions to comprehensive, reliable and relevant spectroscopic data,
in particular
for deformed midshell nuclei where a large diversity of orbitals are accessible,
some of which are involved in the structure of heavier spherical nuclei, i.e., 
placed just above and below the predicted superheavy spherical shell gaps.

The present study of the odd-$Z$ $^{249,251}$Md nuclei is an integral part of this approach by providing
inputs in terms of both proton single-particle and collective
properties.
We report on the previously unobserved ground-state (g.s.) band of $^{251}$Md, assign its single-particle configuration, and deduce the
gyromagnetic factor. 
We also discuss the most intense transition observed in $^{251}$Md using both $\gamma$-ray and conversion-electron spectroscopy, and in $^{249}$Md using $\gamma$-ray spectroscopy alone, as being compatible with octupole correlations.
Finally, a comparison of $^{251}$Md with the $^{255}$Lr nucleus revealed unexpected similarities between transition energies. 
The mechanism leading to these identical bands has been tested with Hartree-Fock-Bogoliubov (HFB) calculations using a Skyrme functional 
and several parametrizations of pairing correlations.

\section{Experimental details}

The experiments were performed at the Accelerator Laboratory of the University of Jyv\"askyl\"a.
The $^{251}$Md nuclei were populated using the fusion-evaporation reaction $^{205}$Tl($^{48}$Ca,$2n$)$^{251}$Md.
The $^{48}$Ca beam was provided at 218\,MeV, resulting in an energy at the middle of the target of 214\,MeV, at which
the fusion-evaporation cross section is about 760\,nb~\cite{Chatillon2007ChT}.
An average beam intensity of $\approx$ 9 particle nA was delivered during $\approx$ 230 h of data taking.
The $^{205}$Tl targets, 99.45 \% enrichment, $\approx$ 300\,$\mu$g/cm$^2$ thick, were sandwiched between a C backing of 20\,$\mu$g/cm$^2$ and
a C protection layer of 10\,$\mu$g/cm$^2$.

\begin{figure*}[htb]
	\begin{center}
		\resizebox{0.98\textwidth}{!}{
			\includegraphics{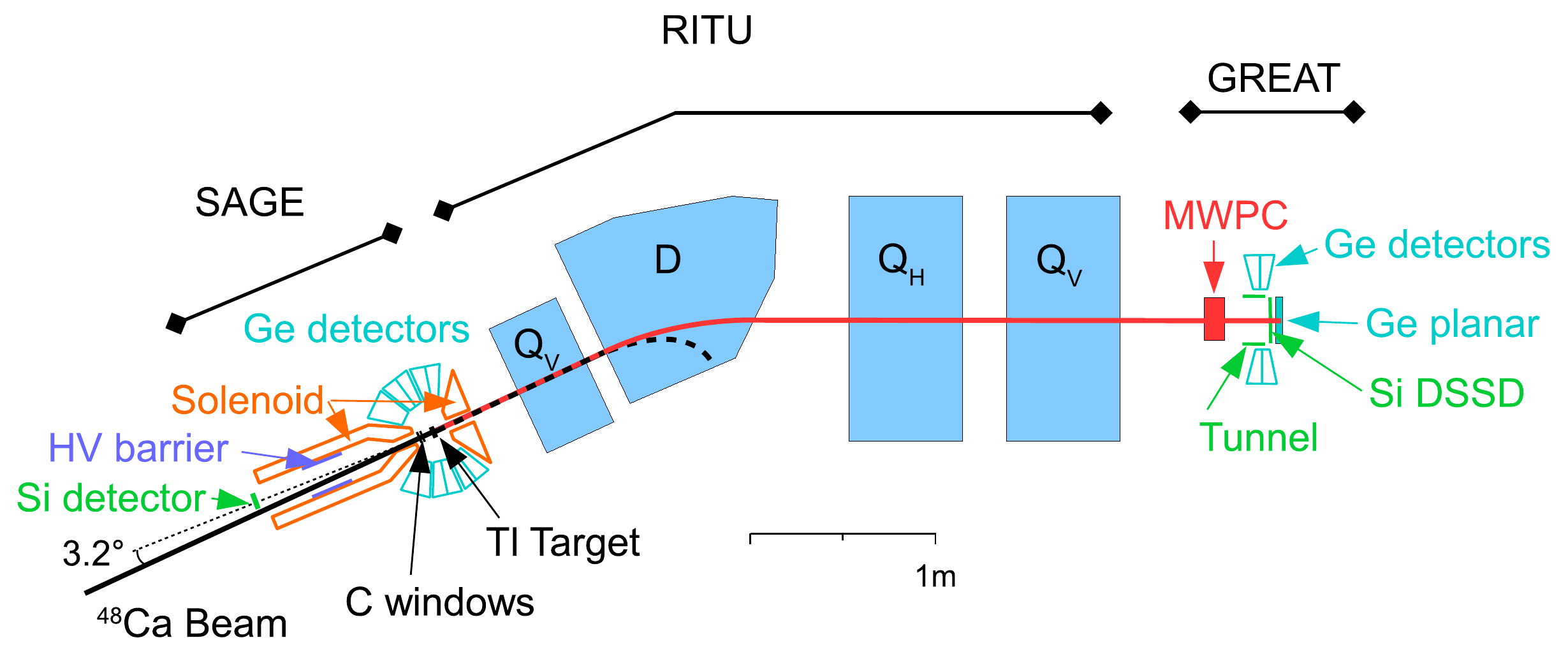} 
		}
	\end{center}
	\caption{Schematic view of the experimental set-up at the Accelerator Laboratory of the University of Jyväskylä, with from left to right the SAGE array for in-beam spectroscopy,  the RITU gas-filled separator, and the GREAT focal-plane detectors. See the text for details.}
	\label{fig:Exp}
\end{figure*}

The experimental  setup is schematically represented in Fig.~\ref{fig:Exp}.
The $^{251}$Md nuclei were separated from the beam and other unwanted products using the Recoil Ion Transport Unit (RITU) gas-filled recoil separator~\cite{Leino1995,Saren2011}
operated at a He gas pressure of 0.4\,mbar.
The recoiling nuclei were detected using the Gamma Recoil Electron Alpha Tagging (GREAT) focal plane spectrometer~\cite{Page2003}.
After passing through a  multiwire proportional chamber (MWPC), ions were implanted in a set of two side-by-side 300-$\mu$m-thick
double-sided silicon strip
detectors (DSSDs). 
Each DSSD had a size of 60$\times$40 mm with 1-mm strips
pitch in both $X$ and $Y$ directions.
The total DSSD counting rate was approximately 250 Hz.
The amplification on the $X$ side was set to a high gain in order to optimize the detection of low-energy conversion electrons in a range of approximately 50-600 keV. 
The $Y$ side was amplified using a lower gain in order to cover energies up to approximately 15 MeV.
Besides the RITU filter, 
an additional selection of the ions of interest was made using a contour gate on the 
the energy-loss $\Delta E$ measured in the MWPC versus the time-of-flight (ToF) measured between the MWPC and the responding DSSD.
The tunnel detectors and the planar and Clover Ge focal-plane detectors were operated during the experiment but not used in the present analysis.
The combined transmission and detection efficiency for the $^{251}$Md residues was estimated at $\approx$ 30\%.
Formal identification using the characteristic $\alpha$ decay (recoil-decay tagging) was not
effective due to the low  $\alpha$-decay branching ratio of 9.5\% for $^{251}$Md~\cite{Chatillon2006}. 
Therefore only the recoil-tagging technique was used, which is adequate since only the reaction channel of interest is open.

Gamma-rays and conversion-electrons emitted at the target position were detected using an array known as Silicon And GErmanium (SAGE)~\cite{Pakarinen2014ChT}:
$\gamma$-rays were detected using Compton-suppressed HPGe detectors (20 coaxial and 24 clovers) having
a total $\gamma$-ray photopeak efficiency of $\approx$ 10\%
at 200\,keV and an average energy resolution of 2.8\,keV FWHM at 1\,MeV.
A stack of 0.5-mm-thick Cu and 0.1-mm-thick Sn absorbers were placed in front of the Ge detectors to reduce the contribution of fission-fragments x-rays. The detection threshold was approximately 20 keV. The maximum counting rate of each coaxial (clover) crystal was kept below $\simeq$ 30 (20) kHz.
After being transported by a solenoid placed upstream the target and tilted 3.2$^{\circ}$ with respect to the beam axis, electrons were detected in a 90-fold segmented Si detector with a thickness of 1\,mm and an active diameter of 48\,mm.
The electron detection efficiency peaks at $\approx$ 6\% for an energy of 120\,keV with an
average energy resolution of 6.5\,keV FWHM in the 50- to 400-keV energy range.
Low-energy atomic electrons were partly suppressed using an electrostatic barrier biased at -35\,kV.
The separation between the He gas-filled region and the upstream beam line, including the electrostatic barrier region, was made using two 50-$\mu$g/cm$^2$ C foils. The maximum counting rate of each segment was kept below 15 kHz. The detection threshold was approximately 30 keV.
Contour gates constraining SAGE time versus the ToF, and SAGE energy versus SAGE time were used to clean the spectra. 
These gates were left wide enough to favor the statistics when using $\gamma$-$\gamma$ coincidences (Secs. \ref{sec:gamma} and \ref{sec:octupole}).
For an analysis that requires intensity measurement, the gates were tightened to favor cleanliness (Secs. \ref{sec:electrons} and ~\ref{sec:profile}).
This can lead, however, to a systematic error in the relative intensities of conversion electrons versus $\gamma$ rays that was estimated at 20 \%.
Digital signal processing was used for the SAGE array (100\,MHz, 14\,bits) while signals from the MWPC and the DSSDs were processed using standard analog electronics and peak sensing analog to digital converters.

The experimental conditions for the study of $^{249}$Md were similar and are detailed in Ref.~\cite{Briselet2019}. 
In brief, the nuclei of interest were produced using the fusion-evaporation reaction $^{203}$Tl($^{48}$Ca,$2n$)$^{249}$Md. 
The $^{203}$Tl targets, 97.08 \% enrichment, $\approx$ 280\,$\mu$g/cm$^2$ thick, were sandwiched between a C backing of 20\,$\mu$g/cm$^2$ and
a C protection layer of 11\,$\mu$g/cm$^2$. 
The $^{48}$Ca beam was delivered at a beam energy of 219 MeV, resulting in an energy of $\simeq$ 215 MeV in the middle of the target. 
Data were taken during $\simeq$ 80 h with a beam intensity of $\simeq$ 13 pnA.
Only $\gamma$ rays were collected.

Data were handled using the triggerless Total Data Readout system~\cite{Lazarus2001} and sorted using the \textsc{grain} software package~\cite{Rahkila2008}.
The theoretical conversion coefficients were calculated using the \textsc{bricc} code~\cite{Kibedi2008}.

\section{Rotational bands in $^{251}$Md}
\subsection{$\gamma$-ray spectroscopy}
\label{sec:gamma}

\begin{figure*}[htb]
\begin{center}
\resizebox{0.98\textwidth}{!}{
  \includegraphics[angle=270]{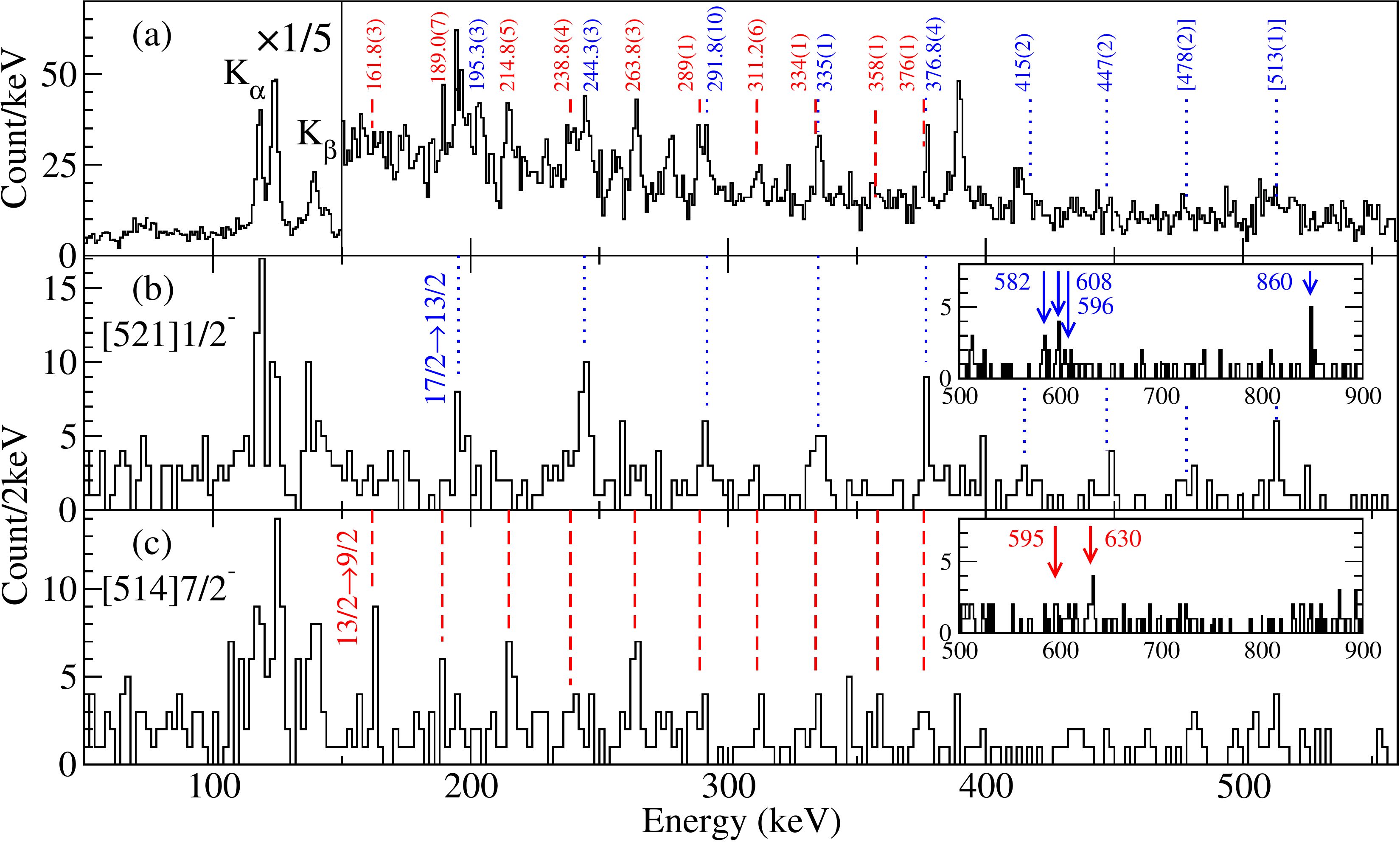} 
}
\end{center}
\vspace{-0.5cm}
\caption{(a) $\gamma$-ray spectra of $^{251}$Md resulting from recoil tagging.
The spectra (b) and (c)  were projected from a sum of gates on the peaks of interest using recoil-gated $\gamma$-$\gamma$ coincidence data.
The transitions in mutual coincidence are shown with dotted blue ($[521] 1/2^-$) and dashed red (previously unobserved band) lines.}
\label{fig:251MdGamma}
\end{figure*}

\begin{figure}[htb]
	\begin{center}
		\resizebox{0.4\textwidth}{!}{
			\includegraphics[angle=270]{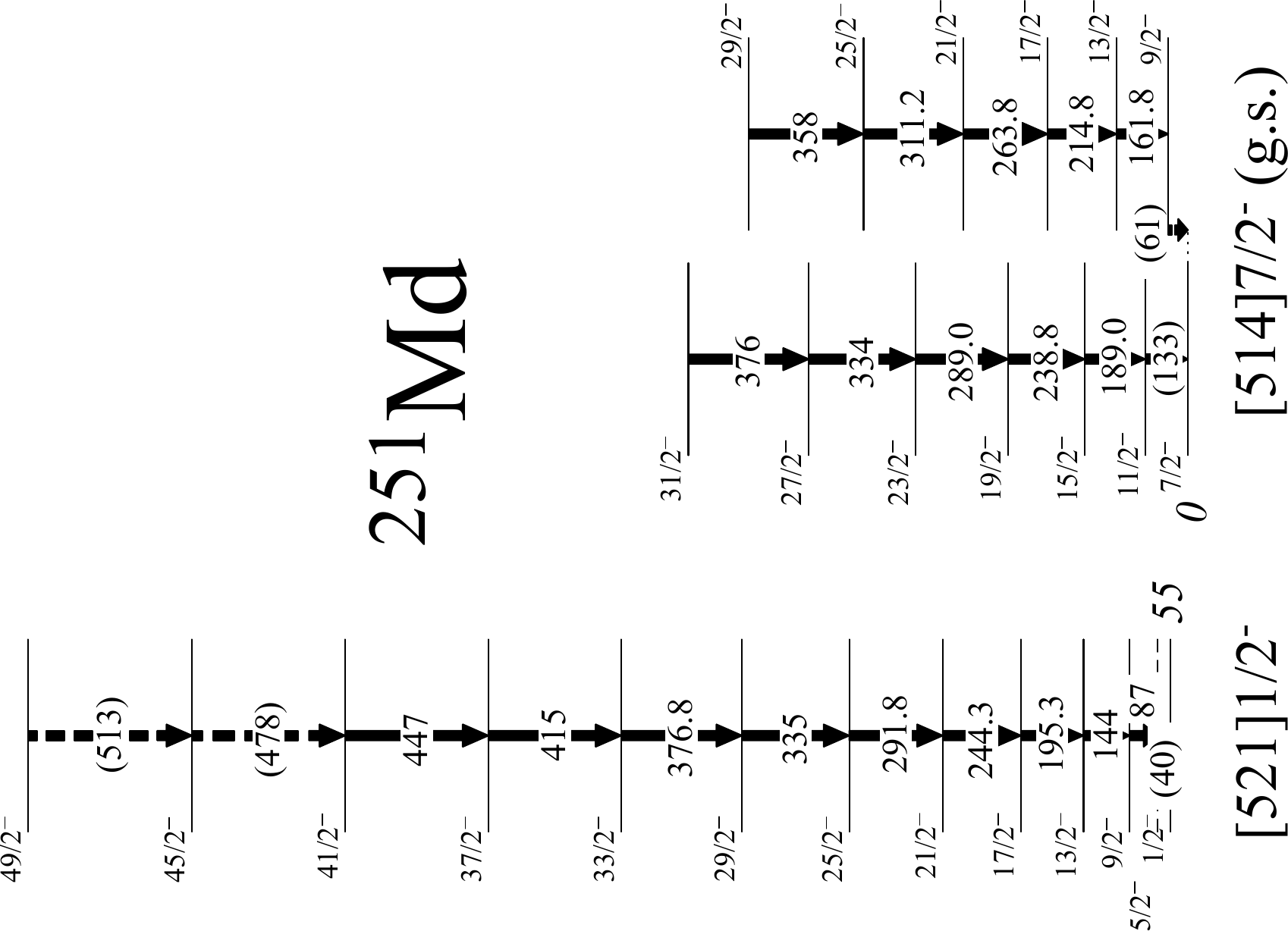}
		}
	\end{center}
	\vspace{-0.5cm}
	\caption{Partial level scheme of $^{251}$Md resulting from this work. 
		All spins and parities are tentative. The band-bead energies are taken from Ref. \cite{Chatillon2006}.
	}
	\label{fig:MdLr}
\end{figure}

The $\gamma$-ray singles spectrum resulting from the recoil-tagging technique is shown in Fig.~\ref{fig:251MdGamma}(a).
The previously observed rotational band structure, interpreted as being built on the proton Nilsson orbital $[521] 1/2^-$~\cite{Chatillon2007ChT}, 
is shown in Fig.~\ref{fig:251MdGamma}(b). 
The spectrum was created from a sum of gates on the peaks of interest, projected from a matrix of recoil-gated $\gamma$-$\gamma$ coincidences.
Compared to the previous work, we cannot confirm the proposed transition at the highest rotation frequency with an energy of 483(1)\,keV for which only six counts were observed~\cite{Chatillon2007ChT}. 
We instead suggest a transition at 478(2)\,keV for which eight counts were collected in the present work.
There is also evidence for an additional transition with an energy of 
513(1)\,keV ($\simeq$ 12 counts).
A structure with a $\gamma$-ray spacing of about half of the former has been found, and is therefore consistent with 
$\Delta I=2$ $E2$ transitions within
the two signature
partners of a rotational structure [Fig.~\ref{fig:251MdGamma}(c)].
Despite the low statistics, each transition resulting from the $\gamma$-$\gamma$ analysis has been found in mutual coincidence with at least
four other transitions in the same band.
It should be noted that the transition at $\simeq$ 334\,keV is a  doublet with the 335\,keV transition of the $K^{\pi}=1/2^-$ band.
For the sake of clarity, a partial level scheme summarizing the results of this work is provided in Fig.~\ref{fig:MdLr}.

The single-particle configurations considered in the following for the previously unobserved band are
those predicted at low energy by the 
macroscopic-microscopic models used in Ref.~\cite{Hessberger2001, Cwiok1994, Parkhomenko2004,Shirikova2013} and  the self-consistent models used in this work (see Sec.~\ref{sec:theory})
as well as those observed by decay spectroscopy in the neighboring Md isotopes~\cite{Hessberger2001,Antalic2009},
which reduces the alternatives to the $[514]7/2^-$ ground-state and to the $[633]7/2^+$ single-particle configurations.

Considering therefore a band-head angular momentum $I=K=7/2$ for the new band,
we can compare the experimental kinematic moment of inertia $\mathcal{J}^{(1)}$
calculated using different angular momenta $I_i \rightarrow I_f$ hypotheses for the transitions (or in other words the number of unobserved transitions),
with the predictions of Chatillon {\sl et al.}~\cite{Chatillon2007ChT}, He {\sl et al.}~\cite{He2009ChT} and Zhang {\sl et al.}~\cite{Zhang2012}.
The best agreement has been obtained using $I_i = 13/2 \rightarrow 9/2$ for the transition at 161.8\,keV.
Extrapolating the $\mathcal{J}^{(1)}$ moment of inertia at lower rotational frequencies yields an energy of 133\,keV for the unobserved
$11/2 \rightarrow 7/2$ transition, and an energy of 61\,keV for the $9/2 \rightarrow 7/2$ one.
The latter is in perfect agreement with the decay spectroscopy of Asai {\sl et al.}~\cite{Asai2015},
which provides an energy of 62\,keV for the first  member of the $7/2^-$ rotational g.s. band.

As shown in the inset of Figs.~\ref{fig:251MdGamma}(b) and~\ref{fig:251MdGamma}(c), the two rotational bands are in coincidence with $\gamma$ rays around 582, 596, 608, and 860\,keV and 595 and 630\,keV, respectively.
Although there are other candidate peaks visible in this region, only those listed here produce coincidences with the rotational band.
In the even-even actinide nuclei, transitions in this energy range are typically observed in the de-excitation of vibrational states,
e.g. $2^-$ states in $^{246}$Cm~\cite{Shirwadkar2019},
$^{250}$Fm~\cite{Greenlees2008}, or $^{252}$No~\cite{Sulignano2012ChT}. 
Also, in the odd-proton $^{255}$Lr, de-excitation of high-$K$ rotational bands proceed via transition in this energy range~\cite{Jeppesen2009ChT}.
However, in our case, coincidences did not allow us to make the link with a collective structure at higher energy.

\subsection{Electron spectroscopy}
\label{sec:electrons}
\begin{figure}[htb]
\begin{center}
\resizebox{0.48\textwidth}{!}{
  \includegraphics{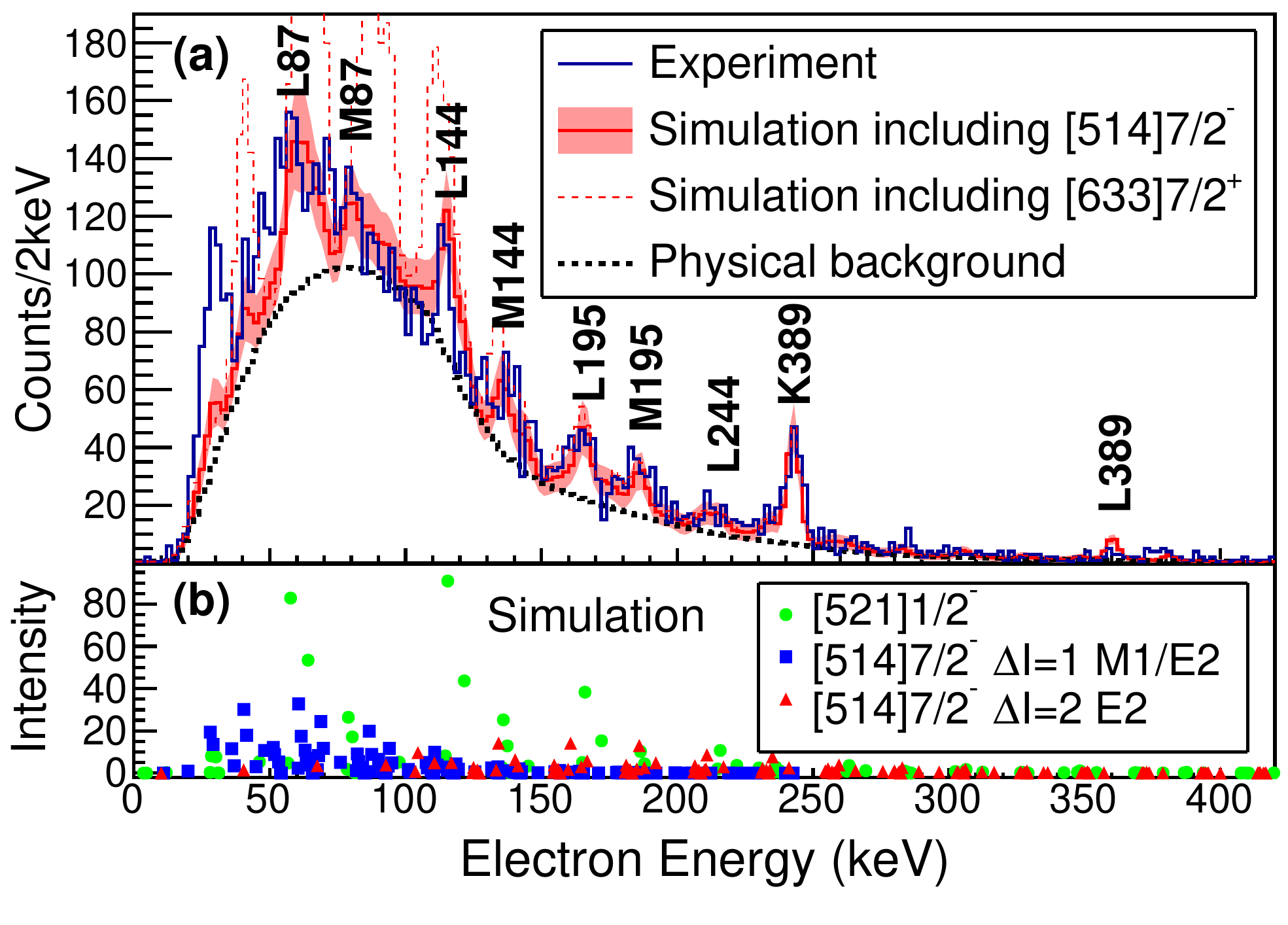} 
}
\end{center}
\vspace{-0.5cm}
\caption{(a)
The experimental $^{251}$Md electron spectrum (solid blue line) is compared to simulations including a physical background (dotted black line) and discrete peaks.
The simulation (solid red line with the propagated uncertainties in shaded red) corresponds to the sum of the transition at 389 keV, the $K^{\pi}=1/2^-$ band, 
the previously unobserved rotational band using the $[514]7/2^-$ configuration and the physical background.
A similar simulation assuming the $[633]7/2^+$ configuration for this band is shown with a dashed red line.
The peaks are labeled with the transition energies preceded by the electronic shell.
(b) The simulated transition intensities corrected according to the experimental detection efficiency
are shown with green circles for the $K^{\pi}=1/2^-$ band.
For the other band, $\Delta I=1$ ($\Delta I=2$) transitions
are shown with blue squares (red triangles) assuming the $[514]7/2^-$ configuration.
}
\label{fig:251MdElectron}
\end{figure}

Turning to in-beam conversion electron spectroscopy, the analysis was based only on the total recoil-gated spectrum due to the paucity of
$\gamma$-electron coincidence data.
The experimental spectrum shown in Fig.~\ref{fig:251MdElectron}(a) was obtained by subtracting the random-correlated background
using
time gates before and after events in prompt coincidence.
L and M components from the 195-keV (L,M195) and 244.3-keV (L244) transitions belonging to the $K^{\pi} = 1/2^-$ band are clearly apparent.
Clearly visible are the L and M conversion lines of a transition at 144 keV,
which fits well with the extrapolated energy for the $13/2^- \rightarrow 9/2^-$ $E2$ transition.
The corresponding $\gamma$ ray overlaps with the $K\beta_2$ x-ray line, which explains why it cannot be evidenced
using $\gamma$-ray spectroscopy alone.
The most intense transition at 389 keV will be discussed further in Sec.~\ref{sec:octupole}.

We also observed peaks below 100 keV with no counterpart in the $\gamma$-ray spectrum.
In order to fully understand the electron spectrum and possibly constrain the single-particle configurations,
we have performed a simulation of the conversion-electron spectra.
We have adopted a purely analytical approach in our simulations. Compared to a Monte Carlo approach, this was possible because the number of ingredients was limited and because the response of the electron detector was quite simple. 
The advantage is that the simulations were fast and that all parameters easily controlled. 
The physics inputs for the rotational bands are described below. 
As soon as the transitions to be simulated have been listed (energy, relative intensity, multipolarity, and mixing ratio), the radiative and converted intensities were calculated using the conversion coefficients. 
The intensity was then corrected from the detector efficiency. 
For the electrons, the spectrum was simply simulated as Gaussian having the experimental resolution, with no background. This simplistic approach is justified by the fact that there is almost no background due to electron (back-)scattering. This was checked using \textsc{nptool}~\cite{Matta2016} (a simulation and analysis framework for low-energy nuclear physics experiments based on \textsc{geant4}~\cite{Agostinelli2003}) in the 50-to 500-keV energy range: More that 85\% of the electrons were indeed fully absorbed in the Si detector. The less than 15\% remaining set of electrons contribute to a background which is mostly concentrated below 150 keV and which resembles the physical background that will be described in the following. 
Since the physical background dominates the majority of the spectrum, this implies that this background is overestimated by about 15\%, which is of the same order of magnitude as the systematic experimental errors. As a matter of fact, the conclusions regarding discrete transitions will not be altered by our assumptions. 
Only the interpretation of the physical background is slightly biased by this hypothesis, but as we will see later, the analysis of this background is essentially qualitative. 
As far as $\gamma$ rays are concerned, the same approach has been chosen, but in this case, a background-less detector response is no longer justified. Simulated $\gamma$-ray spectra are therefore not presented in this study. 
The K x-ray emission after internal conversion was also included in the simulations. The x-ray energies and intensities were taken from Ref.~\cite{Firestone1999}.
The code has been implemented using the root framework~\cite{brun_root_1997}.

The ingredients for the $K^{\pi}=1/2^-$ band were the experimental energies and intensities.
The band was subsequently extrapolated at lower energies using a smooth moment of inertia
resulting in transition energies of 87\,keV ($9/2^- \rightarrow 5/2^-$) and 40\,keV ($5/2^- \rightarrow 1/2^-$), the former being
evidenced in the electron spectrum [Fig.~\ref{fig:251MdElectron}(a)].
These energies are strongly similar to those of bands based on the same single-particle configuration in the
neighboring $^{247}$Bk and $^{251}$Es~\cite{Ahmad1977}.
The transition intensity within the band was deduced from the $\gamma$-ray spectra, corrected for the internal conversion,
and assumed constant for the transitions at 40, 87, and 144\,keV.
The corresponding electron intensities, after taking into account the detector response, are shown with green circles in Fig.~\ref{fig:251MdElectron}(b).

A different approach has been adopted for the other rotational band: In that case, the transition intensities were simulated using
the electromagnetic properties (electric quadrupole and magnetic dipole moments) as an input
assuming either the $[514]7/2^-$ or $[633]7/2^+$ single-particle configuration.
The rotational band was first extrapolated at higher angular momenta using a smooth moment of inertia.
The total experimental intensity profile (converted plus radiative) as a function of the angular momentum, $N(I)$, was fitted using a Fermi distribution
$N(I) = a/\{1 + \textrm{exp}[(I-b)/c]\}$.
This prescription provided the normalization factor $a$,
the average angular momentum entry point $b = 14 \hbar$, and the diffuseness $c = 3 \hbar$.
It is interesting to note that the intensity profile inferred here corresponds remarkably well to that measured
for $^{254}$No at similar conditions of excitation energy
and angular momentum for the compound nucleus~\cite{Henning2014ChT}.
The transition rates of the stretched $E2$ and of the mixed $\Delta I = 1$ $E2/M1$ transitions connecting the two signature partner bands were subsequently calculated from the particle plus rotor model~\cite{Bohr1975}
using the reduced radiative transition probabilities $B(M1)$ and $B(E2)$, and radiative transition rates $T_{\gamma}(M1)$ and $T_{\gamma}(E2)$ :
 \begin{equation}
 \label{equ.bm1}
 B(M1) = \frac{3}{4 \pi} \ ( g_K - g_R )^2 \  K^2 \langle I_i K 1 0 | I_f K \rangle^2 \  [\mu_N^2],
 \end{equation}
 \begin{equation}
 B(E2)  = \frac{5}{16 \pi} \  e^2 Q_{0}^2 \langle I_i K 2 0 | I_f K \rangle^2 \ [e^2 \text{fm}^4], \label{equ.BE2}
 \end{equation}
 \begin{equation}
 T_{\gamma}(M1) = 1.76 \ 10^{13}   E^3 B(M1) \ [\text{s}^{-1}],
 \end{equation}
 \begin{equation}
 T_{\gamma}(E2) = 1.59 \ 10^9      E^5 B(E2) \ [\text{s}^{-1}], \label{equ.TE2}
 \end{equation}
with $g_R \simeq  Z/A$ being the rotational gyromagnetic factor, $\mu_N$ being the Bohr magnetron, and $Q_{0}$ being the electric quadupole moment.
The orbital gyromagnetic factor has been approximated as $g_K = (g_s \Sigma + g_l \Lambda)/K$,
where $g_{s(l)}$ is the nucleon spin (orbital) gyromagnetic factor.
For the protons, we adopted $g_s = 5.59$ and $g_l = 1$.
A reduction of the spin gyromagnetic factor $g_s^{\mbox{eff}} = 0.6  g_s^{\mbox{free}}$ was used, a value generally adopted for heavy nuclei.
The relations above result
in $g_K = 0.66$ and $g_K = 1.34$ for the $[514]7/2^-$ and $[633]7/2^+$ configurations, respectively.
The transition rates, corrected for internal conversion and according to the intensity profile $N(I)$
were subsequently used to calculate the transition intensities along the rotational band.
Actually, the intensity calculations can be performed purely analytically for the entire rotational band.
In practice, the calculations were made from the top of the band, the intensities being calculated for all the transitions steeping downwards.

The resulting simulated electron intensities after taking into account the detector response for the $[514]7/2^-$ hypothesis are shown in Fig.~\ref{fig:251MdElectron}(b)
with red triangles for the stretched $E2$ transitions and with blue squares for $\Delta I=1$ transitions.
As a result, the electron spectrum is dominated by $\Delta I=1$ transitions.
The complexity of the spectrum is obvious along with a fragmented intensity pattern.
However, the simulation of this band essentially generates tails in the peaks of the $K^{\pi}=1/2^-$ band. This is due to two factors: (i) most of the conversion-electron energies of this band are often found close to some of the $K^{\pi}=1/2^-$ band and (ii) the intensities of theses  band transitions is lower than those of the $K^{\pi}=1/2^-$ band.

Experimentally resolving all peaks was not possible given the detector resolution and the collected statistics.
The simulated intensity assuming the $[633]7/2^+$ configuration has similar features but with a larger contribution of
$M1$ transitions (larger $|g_K - g_R|$ value).
Since the total transition intensity was normalized from the $\gamma$-ray spectrum, the resulting electron intensities would exceed those
of the $[514]7/2^-$ case by a factor of $\approx$ 3, which is clearly not compatible with the measurement, and this rules out the $[633]7/2^+$ configuration.

After summing the $[521]1/2^-$ and $[514]7/2^-$ contributions,
there is still a large background in the electron spectrum peaking at $\approx$ 80 keV.
In Fig.~\ref{fig:251MdElectron}(a), a simulation corresponding
to a set of rotational bands with $K$ and the moment of inertia randomly sampled is shown with a dotted black line.
More precisely, each band configuration has been chosen either as a proton 1qp excitation with $ 1/2 \leq K \leq 13/2$
or a single proton excitation coupled to a 2qp, either proton or neutron having
$0 \leq K_{2qp} \leq 8$ and antiparallel spin coupling according to the Gallagher rule~\cite{Gallagher1962ChT}.
The moment of inertia was randomly selected around the value measured for yrast or high-$K$ rotational bands in the Fm-Lr region.
We have arbitrarily taken a fraction of 50\%  proton 1qp excitations, 25\% proton 3qp excitations and
25\% 1qp proton $\otimes$ 2qp neutron configuration.
As shown in Fig.~\ref{fig:251MdElectron}(a), the spectrum resulting from this simulation reproduces the experimental background remarkably well,
with the only adjusted parameter
being its total intensity
scaled to correspond to the experimental spectrum above 250 keV.
It should be noted, however, that the shape of this physical background is largely independent on the assumptions discussed above.
Conversely, it mainly results from a convolution of the electron detection efficiency and the internal conversion coefficients. 
No physics ingredients such as the angular momentum entry point, average gyromagnetic factor, or average moment of inertia notably change the background shape.
This background is, however, fully consistent with rotational bands unresolved experimentally, or from the continuum.
Its intensity can be estimated at a level of 60\% of the total $^{251}$Md population.
It should be reminded here that a background-less response of the electron detector was assumed. As discussed above, this leads to an underestimated simulated background and therefore the present background analysis should be regarded as qualitative.
Our conclusion is, however, entirely consistent with those of Butler {\sl et~al.} in the case of $^{254}$No, for which an electron background has been
well reproduced using a set of $K=8$ rotational bands with 40\% intensity~\cite{Butler2002ChT}.
The decay spectroscopy of $^{254}$No unambiguously confirms the presence of a $K^\pi=3^+$ 2qp state at about 1 MeV and of an isomeric $K^\pi=8^-$ 2qp state at about 1.3 MeV~\cite{Herzberg2006,Tandel2006ChT,Clark2010,Hesberger2010ChT}, on top of which rotational bands were observed.
It was shown in particular in~\cite{Hesberger2010ChT} that the $K^\pi=8^-$ isomeric state 
is fed with an intensity ratio of 28 (2) \%.
Thus, most of the unresolved background identified by Butler {\sl et~al.} arises from the de-excitation of a band based on a high $K^\pi=8^-$ state.
Using in-beam $\gamma$-ray spectroscopy, a rotational band based on a $K^\pi = 8^-$ isomeric state was also observed in $^{250}$Fm~\cite{Greenlees2008} and in $^{252}$No~\cite{Sulignano2012ChT}, with an isomeric ratio of 
37 (2) \% ~\cite{Ketelhut2010ChT} and $\simeq$ 30 \% ~\cite{Sulignano}, respectively. 
A similar situation is thus expected in odd-mass nuclei with the presence of high-$K$ 3qp states.
It should be noted that a high-$K$ isomer has been recently evidenced in $^{251}$Md~\cite{Goigoux}.
It is therefore realistic to interpret the observed electron background
in $^{251}$Md as corresponding to several unresolved bands, built either on high-$K$ 3qp states or on low-lying single-particle configurations such as $[633]7/2^+$ or $[521]3/2^-$.
However, band intensities are expected to be more fragmented compared to even-even nuclei due to the presence of several single-particle states at low excitation energy.

We have also simulated the $\gamma$-ray counterpart of this physical background, which, interestingly,
also reproduces well the $\gamma$-ray background shape in the 150- to 600- keV
energy range.
This must be considered, however, as a qualitative consideration since the response of the Ge detectors was not fully included in our simulations.
With regard to x-rays, their intensity turns out to be very sensitive to $K$, as expected from the strong $K$ dependence of the $B(M1)$ rates [Eq.(\ref{equ.bm1})], but no definitive conclusion can be drawn from the present work.

Finally, the total simulated spectrum ($[521]1/2^-$ and $[514]7/2^-$ configurations, transition at 389 keV plus the physical background)
is shown using a solid red line with the envelope corresponding
to the uncertainty propagation [Fig.~\ref{fig:251MdElectron}(a)].
The global shape is well reproduced except at $\simeq$ 40 keV, 
interpreted as the energy tail of $\delta$ electrons that were not cut by the electrostatic barrier or the electronic threshold.
The same simulation assuming the $[633]7/2^+$ configuration 
is shown with a red dashed line for comparison,
clearly overestimating the measured intensity and again
ruling-out this configuration.

\subsection{Intensity profile}
\label{sec:profile}

\begin{figure}[htb]
\begin{center}
\resizebox{0.5\textwidth}{!}{
  \includegraphics{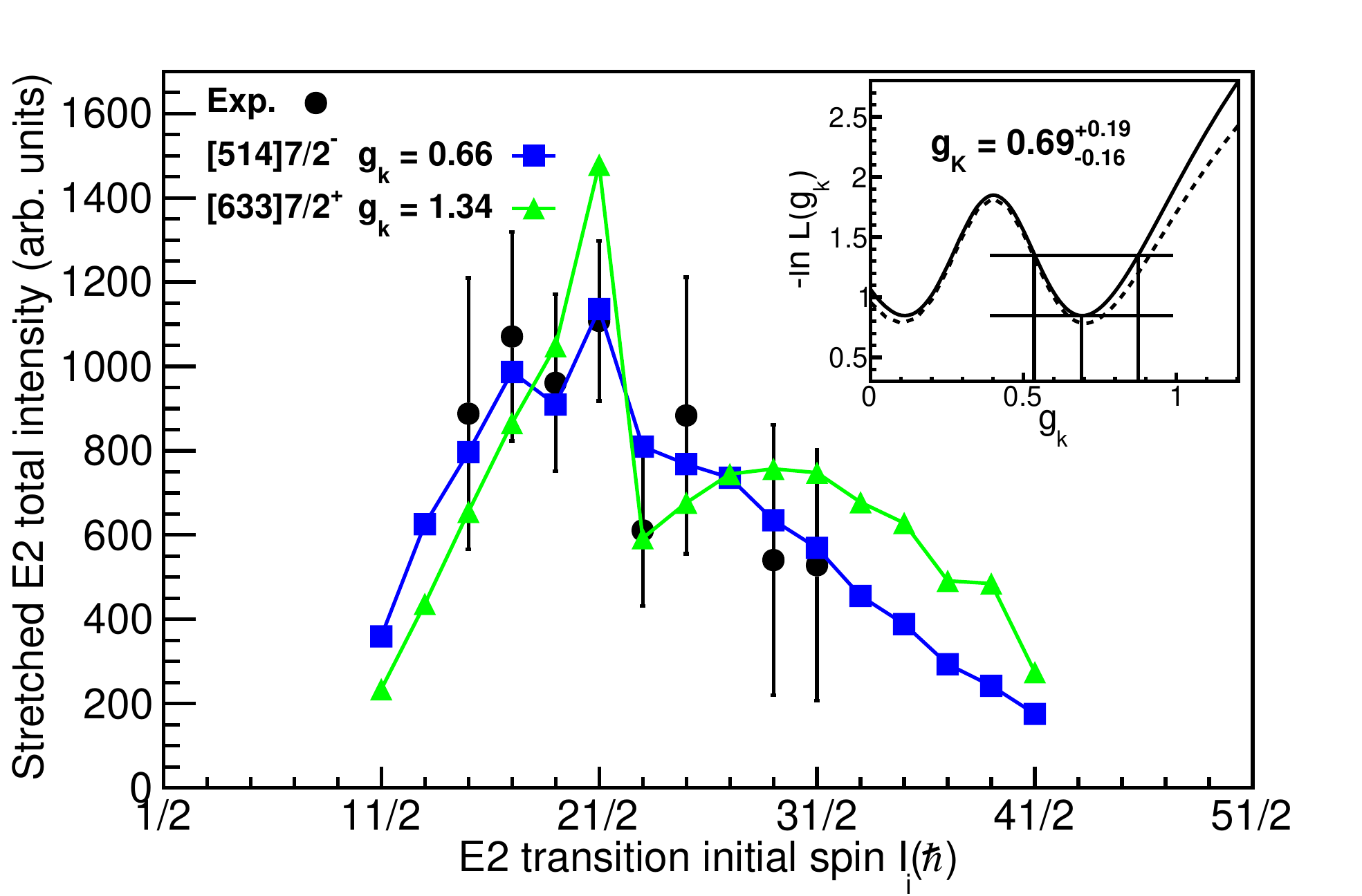} 
}
\end{center}
\vspace{-0.5cm}
\caption{The experimental intensity profile for the $K=7/2$ band as a function of the angular momentum, shown with black dots,
is compared to the band simulation assuming either a gyromagnetic factor of $g_K = 0.66$ (blue squares) or $g_K = 1.34$ (green triangles).
The experimental gyromagnetic factor was deduced using the likelihood estimator $L$, $-\ln L(g_K)$ being plotted in the inset. The solid line was obtained using the entire  experimental profile, the two higher spin transitions being ignored to draw the dashed line.}
\label{fig:profile72}
\end{figure}

The experimental intensity profile of stretched $E2$ transitions for the $K^{\pi}=7/2^-$ g.s. band is shown in Fig.~\ref{fig:profile72}.
The intensities were taken from the $\gamma$ rays in the 189.2-keV ($I_i = 15/2$) to 311.3-keV ($I_i = 25/2$) range, 
corrected from the internal conversion.
Only statistical uncertainties were considered here since a systematic error would simply scale the distribution.
Although associated with large error bars, oscillations around $I_i = 21/2$ are clearly visible.
The intensity profile resulting from the band simulation discussed in Sec.~\ref{sec:electrons} for a gyromagnetic factor
$g_K = 0.66$ is shown with blue squares.

This simulated profile reproduces the experimental data remarkably well, notably the oscillations
and an intensity jump between the states with $I_i = 23/2$ and $21/2$.
Below $I_i = 23/2$, the $\Delta I=1$ transition energies are lower than the K electron-binding energy (145.6 keV)
while they are higher above.
This change results in a sharp decrease of the $M1$ internal conversion coefficient by a factor of $\simeq 5$ and consequently
in a lowering of the
$\Delta I=1$ transition rate below $I_i = 23/2$.
The other impact is an increased stretched $E2$ transition rate below $I_i = 23/2$, clearly visible in the simulation
and evidenced in the experimental data.
For comparison, the intensity profile assuming the $[633]7/2^+$ configuration ($g_K = 1.34$) is shown with green triangles.
Although the jump is still present at the same angular momentum, the intensity profile departs significantly.

The gyromagnetic factor $g_K$ can be deduced using the maximum likelihood technique. 
For convenience, the opposite of the logarithm of the likelihood estimator, $-\ln{L(g_K)}$, is used since  one has simply $-\ln L = 1/2 \chi^2$. This estimator is plotted in the inset of Fig.~\ref{fig:profile72}.  
The estimator, the most likely $g_K$ value and its uncertainty were derived as follows. 
The intensity profile was simulated for different $g_K$ values in the $0 \leq g_K  \leq 1.5 $ range and subsequently compared to the experimental profile using the maximum likelihood technique.
The most likely $g_K$ corresponds to the minimum $-\ln L_{min}$.
The uncertainties were obtained using $-\ln L(g_K + \sigma^+) = -\ln L(g_K - \sigma^-) = -\ln L_{min} + 0.5$, as depicted in the inset of Fig.~\ref{fig:profile72}.
The method allowed us to deduce the most likely gyromagnetic factor $g_K = 0.69^{+0.19}_{-0.16}$.
This value is in remarkable agreement with the value $g_K = 0.66$ estimated above for the $[514]7/2^-$ configuration using the particle-plus-rotor model.
In contrast, for the $[633]7/2^+$ configuration, $g_K = 1.34$ deviates by $2.8 \sigma$ from the most likely value.
It should be noted that the estimator is symmetric with respect to $g_R = Z/A$ and therefore the value $g_K = 0.11$ is still possible, but not favored since it does not correspond to any expected single-particle configuration at low excitation energy.

In order to better estimate the validity of our method, the same test was made taking into account only the first five experimental points. Indeed, the last two points are in the feeding region and therefore more sensitive to the total intensity profile parametrization (a Fermi distribution, as explained above). This analysis lead to $g_K = 0.70^{+0.19}_{-0.16}$, remarkably close to the value obtained using seven experimental points. 
The estimator $-\ln L(g_K)$ using these five experimental values is plotted with a dashed line in the inset of Fig.~\ref{fig:profile72}. 
Both curves are very similar around the most likely value since the last two experimental points have a higher uncertainty and therefore a lower statistical weight.
The second reason is that the model reproduces the two last experimental points very well for the most likely $g_K$ value, contributing  only marginally to the estimator around its minimum. 
In contrast, the two experimental points that contribute most to the estimator are those who delineate the abrupt jump in the intensity profile ($I_i = 23/2$ and 21/2). 
Indeed, this jump amplitude is very sensitive to the gyromagnetic factor, and moreover these two points have a lower statistical uncertainty.

We are therefore confident that our method provides a reliable estimate of the gyromagnetic factor, which fully supports the $[514]7/2^-$ configuration.

\section{Evidence for octupole correlations in $^{249,251}$Md}
\label{sec:octupole}

The most intense $^{251}$Md peak in both $\gamma$-ray and conversion-electron spectra corresponds to a transition at 389 keV. 
Although this transition collects about 16 \% of the de-excitation flow (compared to $\simeq 10 \% $ and  $\simeq 12\%$ for the $K^{\pi}=1/2^-$ and $K^{\pi}=7/2^-$ bands, respectively),
no coincident transition has been observed with a sufficient confidence level. 
Its experimental conversion coefficient
$\alpha_K = 1.84 \pm 0.3 (\textrm{stat}) \pm 0.4 ( \textrm{syst})$ is, within the uncertainties, only compatible with an $M2$ multipolarity, as shown in Table~\ref{tab:conversion}.
We also extracted an L-shell conversion coefficient 
$\alpha_L = 0.75 \pm 0.2 (\textrm{stat}) \pm 0.2 ( \textrm{syst})$.
This value can be considered as an upper limit since the region of the L conversion is not background free.
However,  $\alpha_L$ is again compatible with an $M2$ multipolarity.

\begin{table*}
	\caption{Internal conversion
		coefficient for a transition proceeding via K or L		
		atomic shell with an energy of 389 keV. 
		The theoretical coefficients~\cite{Kibedi2008} for $E1$, $M1$, $E2$, $M3$, $E3$, and $M3$ multipolarities are compared to the experimental value.}
	\label{tab:conversion}	
	\begin{ruledtabular}
		\begin{tabular}{lllllll|l}
			& $E1$ & $M1$ & $E2$ & $M2$ & $E3$ & $M3$ & Exp.\\ 
			\hline 
			$\alpha_K$	& 2.266 10$^{-2}$ & 8.476  10$^{-1}$  & 5.906 10$^{-2}$ & 1.946  & 1.465 10$^{-1}$ & 3.551 & 1.84 $\pm 0.3 \pm 0.4$ \\ 
			\hline 
			$\alpha_L$	& 4.813 10$^{-3}$ &  1.851 10$^{-1}$  & 6.846 10$^{-2}$  & 	6.594 10$^{-1}$ & 5.037 10$^{-1}$ & 2.337 & 0.75 $\pm 0.2 \pm 0.2$ \\ 
			
		\end{tabular} 
	\end{ruledtabular}
\end{table*}

An $M2$ transition of pure single-particle character would have a half-life of $\approx$ 23\,ns and therefore out of the view of the SAGE detection.
A rate enhancement is, however, possible if the initial and final states are coupled via a $2^-$ octupole phonon.
Several examples of octupole-vibration coupling have been observed in the actinide region.
First examples were reported in the 
$^{246,248}$Cm and $^{248,249}$Cf by Yates {\sl et al.} using transfer reactions~\cite{Yates1975ChT,Yates1975}.
A more recent example has been provided in $^{251}$Fm~\cite{Rezynkina2018}.

The occurrence of octupole correlations arise when orbitals with $\Delta l = \Delta j =3$ are close in energy.
In the $Z\sim 100, N \sim 152$ region, this is realized on the neutron side for the $\{[512]5/2^+ (g_{9/2}), [734]9/2^- (j_{15/2})\}$ orbits. 
For the protons, the pair candidates are $\{[512]5/2^- (f_{7/2}), [624]9/2^+ (i_{13/2})\}$ and
$\{[521]3/2^- (f_{7/2}), [633]7/2^+ (i_{13/2})\}$; all these pairs favor a $K^{\pi} = 2^-$ octupole phonon as the lowest collective excitation.
Although the $[521]3/2^-$, $[512]5/2^-$, and $[633]7/2^+$ orbitals have not been observed in  $^{251}$Md yet, the second pair is a better candidate since both orbitals have been predicted at lower energy compared to the first pair~\cite{Parkhomenko2004,Chatillon2006,Shirikova2013}.
In that case, the $3/2^-$ state would have a $[633]7/2^+ \otimes 2^-$ component
and therefore pushed down compared to pure single-particle predictions.

The 389-keV transition can be hence tentatively assigned to a $3/2^- \rightarrow 7/2^+$ transition that could be, because of the coupling with an octupole phonon, of mixed $E3/M2$ character.
We calculated the $\delta^2$ mixing ratio for a $E3/M2$ transition using 
\begin{equation}
\delta^2 (E3/M2) = 
\frac{T_{\gamma}(E3)}{T_{\gamma}(M2)} =
\frac{\alpha_K(M2) - \alpha_K(\mbox{Exp.})}{\alpha_K(\mbox{Exp.})-\alpha_K(E3)}.
\end{equation}
Because of the large experimental uncertainties, our measurement is compatible with no mixing.
However, an upper limit, within one $\sigma$, of $\delta^2 (E3/M2) \le$  0.35 can be provided,
which would nevertheless leave the possibility of a substantial $E3$ component.

\begin{figure}[htb]
	\begin{center}
		\resizebox{0.48\textwidth}{!}{
			\includegraphics{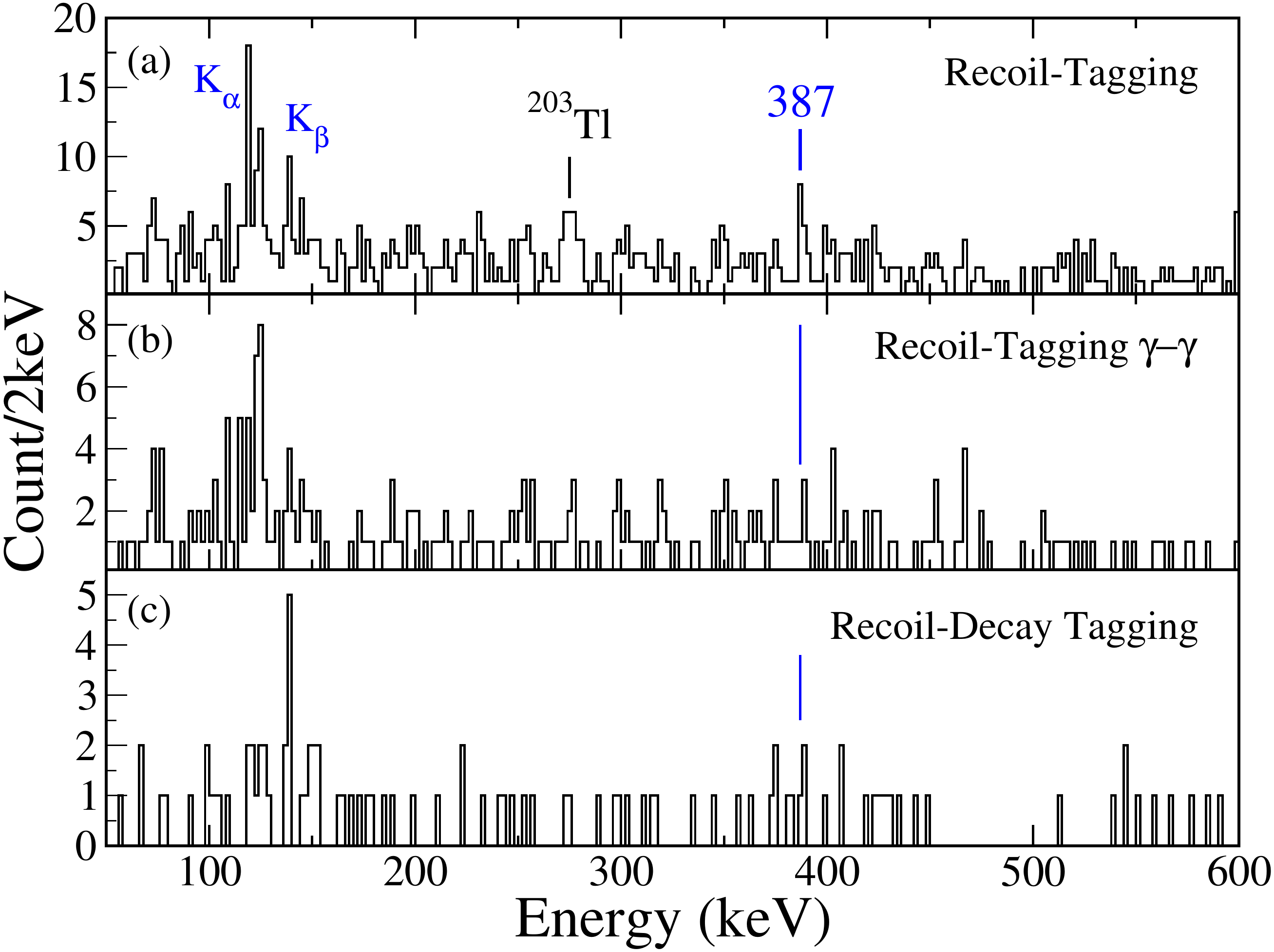}
		}
	\end{center}
	\vspace{-0.5cm}
	\caption{$\gamma$-ray spectra of $^{249}$Md resulting from (a) recoil-tagging,
		(b) sum of $\gamma$-$\gamma$ recoil-tagged coincidences, and
		(c) recoil-decay tagging using a maximum correlation time of 200~s between the recoil implantation and the subsequent $\alpha$ decay. It should be noted that the statistics was not sufficient to establish rotational structures on a solid basis.}
	\label{fig:249MdGamma}
\end{figure}

Interestingly, a transition at 387\,keV has been evidenced in the
neighboring $^{249}$Md as shown in Fig.~\ref{fig:249MdGamma}.
This transition is the most intense in the recoil-tagged spectrum apart from a contamination of $^{203}$Tl Coulex [Fig.~\ref{fig:249MdGamma}(a)]; its intensity decreases using $\gamma$-$\gamma$ coincidences [Fig.~\ref{fig:249MdGamma}(b)], consistent with only few radiative transitions in coincidence as in $^{251}$Md, and finally the transition is still present when tagging on the $^{249}$Md
$\alpha$ decay [Fig.~\ref{fig:249MdGamma}(c)].
The energy and characteristics similar to $^{251}$Md suggests a transition of similar nature in both isotopes.
This is consistent with the fact that several calculations predict a very similar single-particle structure of $^{249}$Md and $^{251}$Md~\cite{Cwiok1994,Parkhomenko2004,Chatillon2006,Shirikova2013}.

\section{Comparison of $^{251}$Md with $^{255}$Lr}
\subsection{Similarity of transition energies: A possible case of identical bands}

An unexpected feature of the $K^{\pi}=7/2^-$ g.s. rotational band of $^{251}$Md is its resemblance to the excited band of $^{255}$Lr~\cite{Ketelhut2009ChT},
both based on the $[514]7/2^-$ orbital.
The transition energies
are indeed identical within the uncertainties up to $I_i = 23/2$, beyond which the difference
slowly increases (Table~\ref{tab:energies} \footnote{It should be noted that no spins were suggested for the transitions in Ref.~\cite{Ketelhut2009ChT}. The spins proposed in the evaluation~\cite{Browne2013} should be excluded since the lowest unobserved transitions, because they were highly converted, were ignored in the evaluated level scheme.}).
On the other hand, rotational bands based on the $[521]1/2^-$ configuration do not exhibit similarities at such a level of precision.
Such a phenomenon of identical rotational bands (IBs)
was first observed in a pair of superdeformed bands (SD) of $^{151}$Tb (first excited band) and $^{152}$Dy (yrast band)~\cite{Byrski1990}, later confirmed in numerous cases
and which turned out to be emblematic of SD bands. 
The phenomenon of IBs has also been observed at intermediate and normal deformations, with bands having a variable degree of similarity, and sometimes with IBs for nuclei that differ substantially in mass (Ref.~\cite{Baktash1995} and references therein).

\begin{table}
\caption{Rotational band transition energies (keV) and initial angular momentum $I_i$ for $^{251}$Md (this work) and $^{255}$Lr~\cite{Ketelhut2009ChT}.
Tentative transitions are written in brackets.}
\label{tab:energies}
\begin{ruledtabular}
\begin{tabular}{lll|lll}
\multicolumn{3}{c|}{$[514]7/2^-$} & \multicolumn{3}{c}{$[521]1/2^-$} \\
$I_i (\hbar)$ & $^{251}$Md & $^{255}$Lr & $I_i (\hbar)$ & $^{251}$Md & $^{255}$Lr \\
\noalign{\smallskip}\hline\noalign{\smallskip}
13/2 & 161.8 (3) &           & 9/2  & $[87]$  & 		     	\\
15/2 & 189.0 (7) & 189 (1)   & 13/2 & 144 (1) & 		     	\\
17/2 & 214.8 (5) & 215 (1)   & 17/2 & 195.3 (3)    & 196.6 (5)   \\
19/2 & 238.8 (4) & 239 (1)   & 21/2 & 244.3 (3)    & 247.2 (5)  \\
21/2 & 263.8 (3) & 264.6 (5) & 25/2 & 291.8 (10)   & 296.2 (5) \\
23/2 & 289   (1) & 288.4 (5) & 29/2 & 335   (1)    & 342.9 (5) \\
25/2 & 311.2 (6) & 314.0 (5) & 33/2 & 376.8 (4)    & 387   (1) \\
27/2 & 334   (1) & 338 (1)   & 37/2 & 415   (2)    & 430 (1) \\
29/2 & 358   (1) & 359 (1)   & 41/2 & 447   (2)    & \\
31/2 & 376   (1) & 384 (1)   & 45/2 & [478 (2)]    & \\
33/2 &           &           & 49/2 & [513 (1)]    & \\          
\end{tabular}
\end{ruledtabular}
\end{table}

It is worth reminding that, at moderate deformation, the classical moment of inertia 
of a rigid homogeneous body is proportional to $A^{5/3}(1+0.31\beta)$  \textit{}
\cite{Rowe1970}. Phrased differently, the transition energies of such rotating 
rigid bodies scale with $A^{-5/3}$ for the same deformation. Therefore, for the $^{251}$Md-$^{255}$Lr 
pair, an energy difference of $\simeq$ 3\% between the transition energies of the two bands is 
expected in such  a
purely macroscopic framework, significantly higher than the observation.
For the pair of bands based on the 7/2$^-$ single-particle state, five transitions are identical within one keV, which could be considered as not very impressive at first sight compared to e.g. the phenomenon in SD bands.
As we will discuss in the following, however, this case turns out to be unique in the transuranium region.
It is also interesting to note that the transition energies for the $K^{\pi} = 1/2^-$ band are larger in $^{255}$Lr than in $^{251}$Md, while the opposite effect is expected according to the  $A^{-5/3}$ scaling of rotational energies at fixed angular momentum and deformation.

Numerous mechanisms have been advocated to explain the IB phenomenon
such as (i)  the spin alignment of specific orbitals along the rotation axis in the strong coupling 
limit of the particle-rotor model,
(ii) the role of symmetries and in particular the pseudo-SU(3) scheme,
(iii)  the role of orbitals not sensitive to the rotation, in particular, those having a high density in the equatorial plane (low
number of nodes $n_z$ in the plane perpendicular to the symmetry axis),
(iv) the role of time-odd terms, etc.; see Ref.~\cite{Baktash1995} and references therein.
 None of them were fully satisfactory since they were
neither predictive nor capable of identifying the underlying mechanism.
Some global analyses using mean-field approaches suggest that the mechanism is not as simple as a quantum alignment or purely related
to single-particle properties, but results from a cancellation of several contributions (deformation, mass, pairing),
resulting in the identical bands (e.g. Refs.~\cite{Ragnarsson1990, Ragnarsson1991,Szymanski1995,Satula1996}).

Identical bands were previously reported in even-even transuranium nuclei~\cite{Ahmad1991}: The three or four first  ground-state band transitions in  $^{240}$Pu, $^{244,246}$Cm, and $^{250}$Cf are identical within 2 keV. 
The more recent improved spectroscopy of $^{240}$Pu~\cite{Wang2009ChT}, $^{246}$Cm and $^{250}$Cf~\cite{Hota2012ChT} has shown that the transition energies deviate significantly above $I_f$ = 8.
More impressive are the ground-state bands of $^{236,238}$U that are identical up to spin $I_f =22^+$ within 2 keV~\cite{Ahmad1991}. 
In this reference, this
has been interpreted in this region of midshell nuclei as the filling of orbitals driving small deformation changes that counteract the mass dependence.
In any event, even if these bands cannot all be qualified as being identical, these 
cases recall that the systematics of moments of inertia can locally deviate very strongly from the 
overall scaling with  $A^{5/3}$ \cite{Rowe1970}.

To establish whether other identical bands are present in the transuranium region, we
have inspected all bands having at least eight measured transitions, which represent 30 cases in even-even nuclei and 29 bands in odd-$N$ or -$Z$ nuclei (odd-odd nuclei were not considered). 
The data were taken from the Evaluated Nuclear Structure Data File (ENSDF) and from more recent publications or unpublished works~\cite{Hota2012ChT,Birkenbach2015ChT,Ketelhut2010ChT}.
This survey revealed two other even-even pairs that are identical within 2 keV for the four lowest spin transitions: They are respectively $^{236}$Pu and $^{242}$Pu, 
and $^{250}$Fm and $^{254}$No.
The equality of the transitions, however, is verified over a few
transitions only.
We also mention the case of the ($^{246}$Pu-$^{252}$No) pair whose transition energies are identical within 2 keV up to $I_f$ = 6. 
This case can be hardly explained as these nuclei differ by eight protons and two neutrons, and may be considered as an accidental degeneracy.
It is worth mentioning that the general trend of the $2^+$ collective state in the $N=152$, $Z=100$ region can be well explained by a change of the moment of inertia because of a 
reduction or increase in pairing correlations
when approaching or leaving the
deformed shell gaps (see the discussion, e.g. in Ref.~\cite{Theisen2015}).
Several nuclei, however, deviate from this trend (in particular the Cf isotopes).
No explanation that would convincingly explain all cases has been found so far.
Also, octupole correlations have been evidenced in the $^{240-244}$Pu~\cite{Wang2009ChT,Wiedenhoever1999ChT} isotopes.
Clearly, beyond-mean field effects have to be taken into account in that case.
Except the $^{251}$Md-$^{255}$Lr pair, we have not identified other cases of odd-mass nuclei having identical transitions between them, or having identical transitions with one of their neighboring even-even nuclei.

An intriguing fact in the IBs discussed here is that the nuclei differ by four mass units, more precisely an $\alpha$ particle.
In the rare-earth-metal region, an $\alpha$ chain of even-even nuclei with bands identical  up to $I_f =12$ has been identified: $^{156}$Dy, $^{160}$Er, $^{164}$Yb, $^{168}$Hf, $^{180}$Os (and $^{172}$W to a lesser extent)~\cite{Casten1992}.
An interpretation based on the algebraic interacting boson model was proposed in the same reference.
The quadrupole moment is linked to the $N_p N_n$ product, $N_p$ ($N_n$) being the number of valence protons (neutrons) with respect to the nearest magic shell. 
This product is similar for all these nuclei and moreover
in the language of the interacting
boson model, these nuclei form a $F$-spin multiplet having the same number of valence particles $N_p + N_n$; i.e., they are predicted to have a similar structure.
However, the major drawback of this approach is that it overpredicts the occurrence of identical bands.
Deficiencies to reproduce the gyromagnetic factor for these nuclei have also been noticed~\cite{Stuchbery1995}.

\begin{figure}[htb]
	\begin{center}
		\resizebox{0.48 \textwidth}{!}{
			\includegraphics{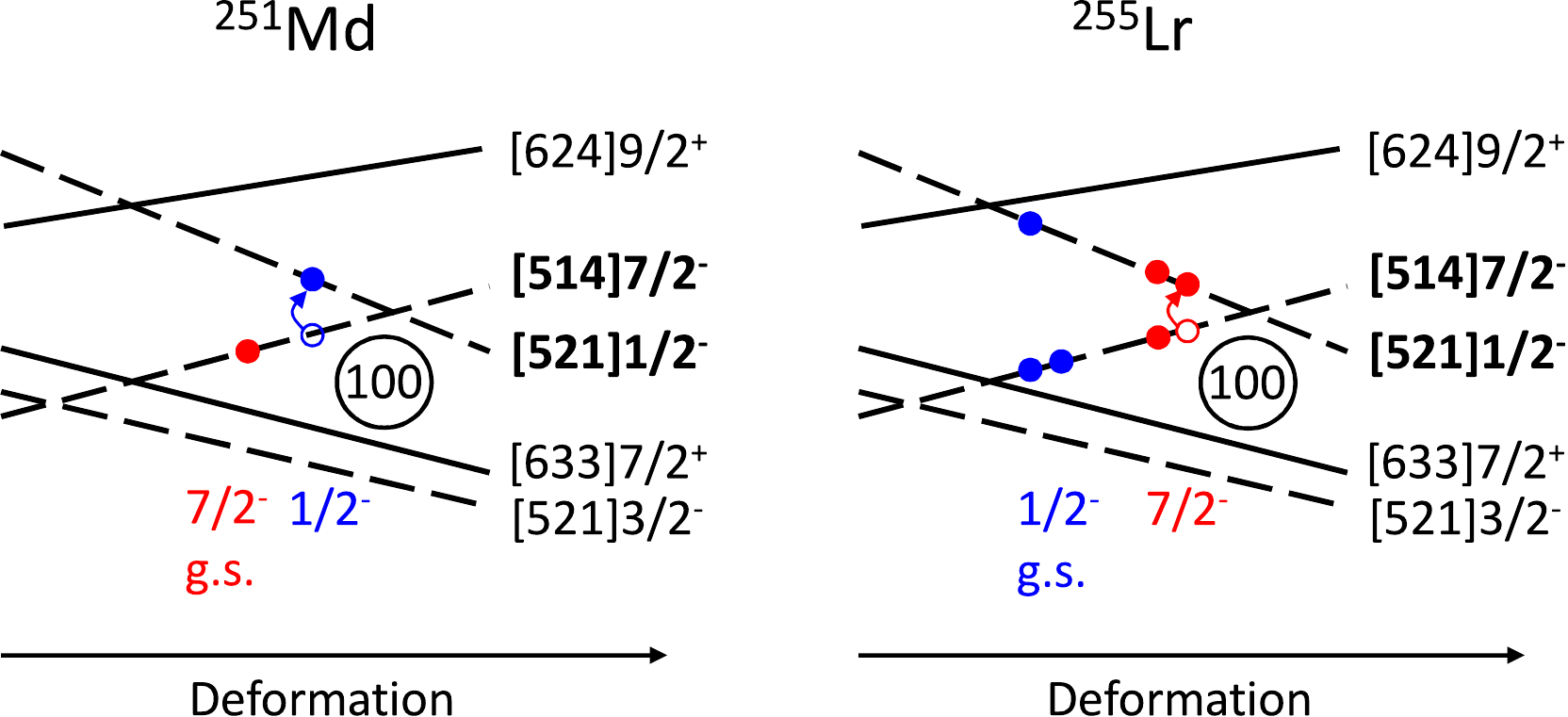} 
		}
	\end{center}
	\caption{Schematic proton Nilsson diagram for $^{251}$Md and $^{255}$Lr inspired from calculations using a Woods-Saxon potential (see, e.g. Ref.~\cite{Chasman1977}).
	}
	\label{fig:orbitals}
\end{figure}

The nuclear structure and deformation 
can be expected to 
change when filling deformation-driving orbitals (either down- or up- sloping as a function of the deformation driving the nucleus toward larger or lower deformation, respectively).
In this respect, it is interesting to note that because of the 
sequential filling of proton levels, the ground-state and the first excited state configurations 
of $^{251}$Md versus $^{255}$Lr are interchanged. Depending on its characteristics, the additional 
occupied pair of orbits in $^{255}$Lr can lead to subtle deformation changes 
between the two nuclei; see Fig.~\ref{fig:orbitals}.

In the ground state of $^{251}$Md, one level out of the pair of 
up-sloping $[514]7/2^-$ orbitals is filled, while in $^{255}$Lr the $7/2^-$ state 
corresponds to the same configuration plus a pair of filled $[521]1/2^-$ down-sloping 
orbitals, the latter driving the nucleus toward slightly larger deformations.
The deformation-driving effect in $^{255}$Lr for this configuration goes therefore in the 
same direction as the overall $A^{5/3}$ macroscopic dependence of the rigid moment 
of inertia. 
The filling of orbitals as such therefore cannot explain the experimental finding 
of identical $7/2^-$ bands, quite on the contrary: Based on this simple argument, the 
two bands should be even more different than can be expected from the global scaling
of moments of inertia.
On the other hand, the $[514]7/2^-$ pair is filled for the $1/2^-$ $^{255}$Lr ground state, 
therefore driving the nucleus towards a lower deformation compared to the $^{251}$Md 
$1/2^-$ excited state, which goes in the opposite direction as the $A^{5/3}$ trend: 
A mechanism consistent with identical bands for the $[521]1/2^-$ configuration, 
again in contradiction with the experimental finding.

Furthermore, there is a pair of neutron orbits that is filled to pass from $^{251}$Md to 
$^{255}$Lr, namely in the $[734]9/2^-$ Nilsson orbital. 
According to calculations using a Woods-Saxon potential (see, e.g.~\cite{Chasman1977})
or calculations presented below (see the self-consistent Nilsson diagram in Fig.~\ref{fig:nilsson:n}), this level is not sloping around the ground-state deformation, 
which justifies ignoring neutron levels at the present level of discussion.

There is also the experimental observation to consider that the transition energies for the 
 $K^{\pi} = 1/2^-$ bands change in the opposite direction to that expected from the 
$A^{-5/3}$ scaling.
Moreover, from a purely macroscopic point of view, the deformation of $^{255}$Lr should  decrease from $\beta \simeq 0.3$ to $\simeq$ 0.21 to compensate for the mass difference ($A^{5/3}$ term)
and lead to the same energies as $^{251}$Md.
Therefore, the mass-deformation compensation mechanism discussed above 
does not have the correct order of magnitude since only small deformation changes are expected and
cannot explain simultaneously the larger transition energies for the $K^{\pi} = 1/2^-$  $^{255}$Lr band and 
IBs for the $K^{\pi} = 7/2^-$ bands, unless one assumes that there is an additional mechanism that decreases the moment of inertia in $^{255}$Lr. If the mechanism is the same for both configurations, 
then it just has the right size for the $7/2^-$ bands to make them identical, but ``overshoots" for the $1/2^-$ bands. 

The mass-deformation compensation mechanism resulting from the filling of levels
is therefore unable to explain the experimental findings.
There clearly have to be additional
compensation effects, for example, from changes in pairing correlations or the alignment
of single-particle states as proposed in Refs. \cite{Ahmad1991,Ragnarsson1990,Espino1989,Stephens1990,Zhang1992}  for the
observation of identical bands found for pairs of rare-earth-metal nuclei.

In this respect, it is interesting to note that the $^{251}$Md and
$^{255}$Lr nuclei are the neighbors of $^{250}$Fm and $^{254}$No
respectively with one additional proton. As already mentioned,
the yrast bands of these two even-even nuclei are also identical
for the first four transitions.
For the same reasons discussed above for the case of $^{251}$Md and
$^{255}$Lr, the mass-deformation compensation mechanism also cannot explain the similarity between the yrast band of $^{250}$Fm and
$^{254}$No. With $^{250}$Fm being proton magic deformed and
$^{254}$No neutron magic deformed, a simple explanation of the change
in moment of inertia in terms of a change in pairing correlations
is also not straightforward.
The rotational $K^\pi = 1/2^-$ bands of $^{251}$Md and $^{255}$Lr
can be phenomenologically described by the coupling of a proton
in the $K^\pi = 1/2^-$ orbit to the ground-state band of $^{250}$Fm
and $^{254}$No, respectively. In the most basic version of such a
model \cite{Rowe1970}, one automatically obtains identical 
$K^\pi = 1/2^-$ bands in $^{251}$Md and $^{255}$Lr as well.
However, this is not observed, indicating that there are additional
changes that are not the same when passing from $^{250}$Fm to
$^{251}$Md and from $^{254}$No to $^{255}$Lr, respectively.
For the $K^\pi = 7/2^-$ bands the situation is even more complicated
since for $^{251}$Md the $K^\pi = 7/2^-$ band could be interpreted
as a proton in the $K^\pi = 7/2^-$ orbit coupled to the ground-state
band of $^{250}$Fm, whereas for $^{255}$Lr the $K^\pi = 7/2^-$
band corresponds a 2p-1h excitation relative to the ground-state
band of $^{254}$No.

\subsection{Self-consistent mean-field analysis}
\label{sec:theory}

To better understand the conditions for the emergence of identical bands for the nuclei studied 
here, we performed microscopic
cranked self-consistent-mean-field calculations  for the $K^\pi = 1/2^-$ and 
$K^{\pi} = 7/2^-$ bands in $^{249}$Md, $^{251}$Md, and $^{255}$Lr. The calculations were made 
with the coordinate-space solver \textsf{MOCCa} \cite{Ryssens2016,MOCCa} that is based on 
the same principles as the code used for the Skyrme-HFB calculations reported in Refs.~\cite{Chatillon2006,Chatillon2007ChT,Ketelhut2009ChT,Dobaczewski2015ChT}. We employ the 
recent SLy5s1 parametrization of the Skyrme energy density functional (EDF) \cite{Jodon2016} that 
was adjusted along similar lines as  the widely-used SLy4 parametrization \cite{Chabanat1998a} 
used in those references, but with a few differences in detail, the most important one being a 
constraint on the surface  energy coefficient that leads to a much better description of fission 
barriers of heavy nuclei \cite{Rysssens2019a}. As pairing interaction we choose a so-called 
``surface pairing" with cutoffs as defined in Ref.~\cite{Rigollet1999}. 

\begin{figure}[htb]
	\begin{center}
		\resizebox{0.48 \textwidth}{!}{
			\includegraphics{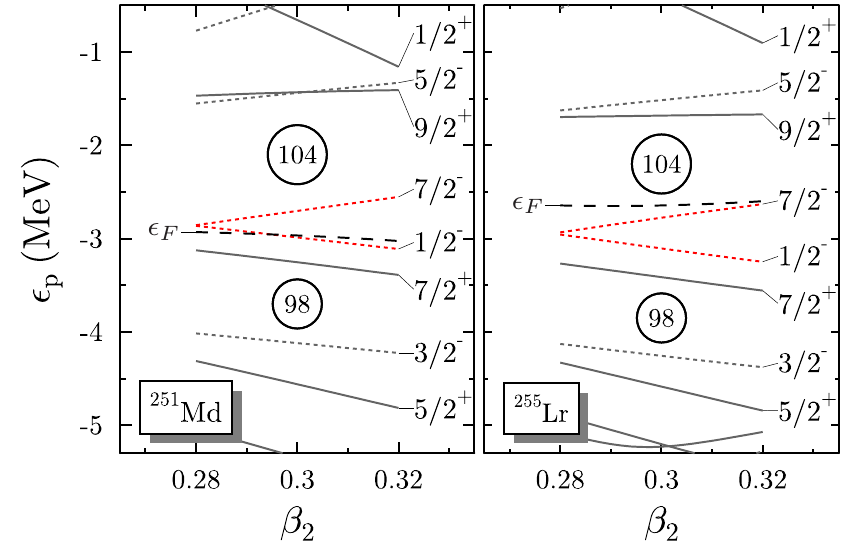}
		}
	\end{center}
	\caption{Nilsson diagram of proton single-particle levels around the Fermi energy for mass 
	quadrupole deformations $\beta_2$ as defined in Ref.~\cite{Rysssens2019a} around those of the 
	ground state for false vacua of $^{251}$Md and $^{255}$Lr, calculated with SLy5s1 and 
	stabilized HFB pairing. The $K^{\pi}=1/2^-$ and $K^{\pi}=7/2^-$ levels are highlighted in color. 
	The Fermi energy $\epsilon_F$ is indicated by a dashed line in each panel.
	}
	\label{fig:nilsson:p}
\end{figure}

\begin{figure}[htb]
	\begin{center}
		\resizebox{0.48 \textwidth}{!}{
			\includegraphics{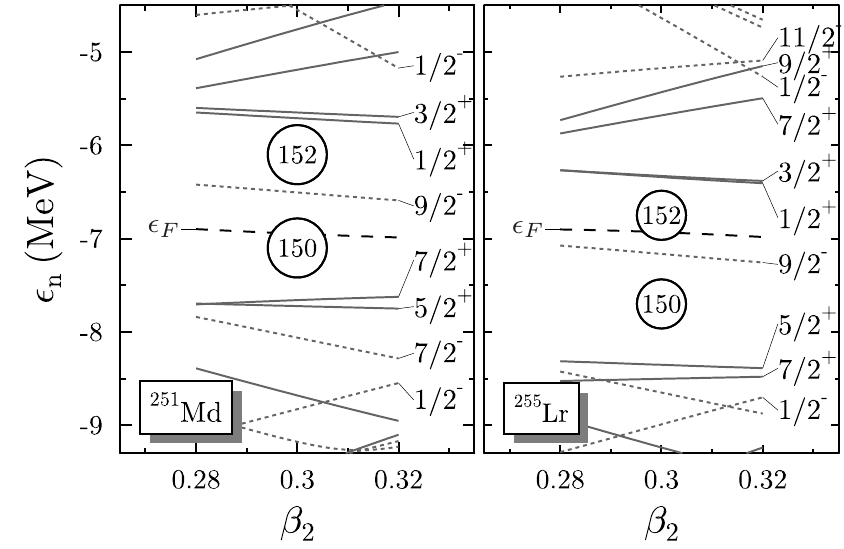} 
		}
	\end{center}
	\caption{Same as Fig.~\ref{fig:nilsson:p}, but for neutrons.}
	\label{fig:nilsson:n}
\end{figure}

For further discussion, it is important to recall that SLy5s1 does not reproduce the empirical
deformed shell closures at $Z=100$ and $N=152$ \cite{Theisen2015,Asai2015,Ackermann2017}, 
a property that it shares with SLy4 and almost all other available nuclear EDFs that have been 
applied to the spectroscopy of very heavy nuclei so far  \cite{Dobaczewski2015ChT,Bender2013}. 
Instead, SLy5s1  gives prominent deformed proton gaps at $Z=98$ and $Z=104$, and an 
additional deformed neutron gap at $N=150$; see Figs.~\ref{fig:nilsson:p} and~\ref{fig:nilsson:n}. 
For the Md and Lr isotopes discussed here, the Fermi energy is in the direct vicinity of these shell 
closures, which has some influence on the calculated properties of their rotational bands.

We observe that, for a given Skyrme interaction, the similarity of in-band 
transition energies for different nuclei depends sensitively on the details
of the treatment of pairing correlations.
To illustrate this finding, four different options will be compared. The first one 
is the HFB + Lipkin-Nogami (HFB+LN) scheme as defined in Ref.~\cite{Rigollet1999} with a pairing 
strength of $-1250 \, \text{MeV} \, \text{fm}^3$ for protons and neutrons that was adjusted to 
describe superdeformed rotational bands in the neutron-deficient Pb region. This prescription has
 also been used in Refs.~\cite{Chatillon2006,Chatillon2007ChT,Ketelhut2009ChT}. 
 The second option is a HFB+LN scheme with a reduced pairing strength of
 $-1014 \, \text{MeV} \, \text{fm}^3$ that was adjusted to reproduce the kinematic moment of 
inertia of $^{252}$Fm at low spin when used with SLy5s1 \cite{BendertbpSHEbands}. 
While the LN scheme is a popular prescription to avoid the breakdown of HFB pairing correlations 
in the weak-pairing limit, it is known to have some conceptual problems, the most prominent 
one not being variational. As an alternative, Erler \textit{et~al.} \cite{Erler2008} proposed a small 
modification that can be applied to any pairing interaction and that prevents the breakdown of pairing
when being inserted into a standard HFB calculation. Their fully variational stabilized HFB scheme 
was used as a third pairing option, again with a surface pairing interaction of strength 
$-1250 \, \text{MeV} \, \text{fm}^3$. 
As the fourth option we use the standard HFB scheme as the most basic reference case, again 
with a pairing strength of $-1250 \, \text{MeV} \, \text{fm}^3$.

Independent of the pairing option chosen, we find a calculated $1/2^-$ ground state for $^{249}$Md 
and $^{251}$Md, but a $7/2^-$ ground state for $^{255}$Lr. In each case, the other state is a 
low-lying excitation at less than 160 keV. This result is at variance with experimental data, for 
which the relative order of these levels is the other way round \cite{Chatillon2006}. This finding is 
intimately connected to the incorrect deformed gaps found 
in the Nilsson diagram of Fig.~\ref{fig:nilsson:p}: in order to obtain the correct level sequence, the 
$K^{\pi} = 1/2^-$ level has to be pushed up relative to the other levels such that it is above the 
$K^{\pi} = 7/2^-$ level at all relevant deformations. This would open up a gap at $Z=100$ and 
significantly reduce the $Z=104$ gap; see the detailed discussion of this point in Ref.~\cite{Chatillon2006}.
Similar problems for the relative position of these two levels were found for virtually all widely used nuclear EDFs \cite{Dobaczewski2015ChT,Bender2013}.
It is noteworthy that the UNIDEF1$^\text{SO}$ parametrization of Ref.~\cite{Shi2014}
for which the spin-orbit interaction has been fine-tuned to give deformed 
$Z=100$ and $N=152$ shell gaps does not improve on the relative position of these
two levels. In addition, it predicts that the $9/2^+$[624] level is nearly
degenerate with them, which is difficult to reconcile with the systematics of
band heads in this region.

The Nilsson diagrams of Figs.~\ref{fig:nilsson:p} and~\ref{fig:nilsson:n} have been calculated for
false vacua, 
meaning HFB states 
that have the correct  
odd particle number on average,
but no blocked quasiparticles.
It is noteworthy that the relative positions of many neutron and proton levels visibly change when 
going from $^{251}$Md to $^{255}$Lr: Filling a further pair of neutron and proton orbits changes
all other levels through self-consistency. Such self-consistent rearrangement of deformed shells
seems to be a general feature of heavy deformed nuclei when calculated within self-consistent 
models \cite{Bender2013}.

\begin{figure}[htb]
	\begin{center}
		\resizebox{0.48 \textwidth}{!}{
			\includegraphics{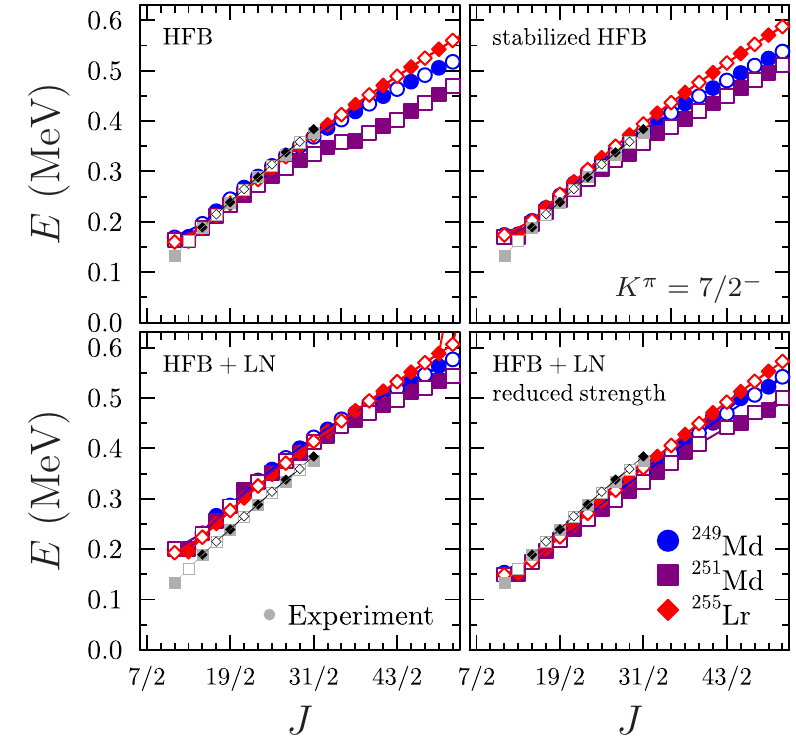} 
		}
	\end{center}
	\caption{$E2$ transition energies in the $K^\pi = 7/2^-$ band
	calculated with SLy5s1 and the pairing options for the three nuclei as indicated. 
	Calculated values are plotted in color as indicated, whereas 
	experimental values are plotted with smaller gray and black symbols for $^{251}$Md and $^{255}$Lr, respectively.
	Full symbols indicate transitions in the favored band, and open symbols indicate transitions in the 
	nonfavored band. 
}
\label{fig:egamma:35-}
\end{figure}

The rotational levels in each band have been constructed by solving the cranked HFB equations 
with a constraint on the collective angular momentum $I_z = \langle \hat{J}_z \rangle$ such that 
$J(J+1) = I_z^2 + K^2$, with $K$ held fixed at $7/2$ or $1/2$, respectively. 
The odd particle can be put either into the orbit 
with $+K$ or $-K$, which leads to two different solutions of the HFB equations that we identify
with the states in the two signature-partner bands that can be observed experimentally
\cite{deVoigt1983}. 
With increasing spin $I_z$, 
one finds
a signature splitting 
between the two calculated partner bands into an energetically favored and non-favored band. 
For the calculated and observed $K^\pi = 7/2^-$ bands, the signature splitting is too small to be resolved 
on the  plots. For the calculated $K^\pi = 1/2^-$ band, however, it is quite substantial. As there are no 
experimental data for the non-favored $K^\pi = 1/2^-$ band, we will not discuss its properties here.

\begin{figure}[htb]
	\begin{center}
		\resizebox{0.48 \textwidth}{!}{
			\includegraphics{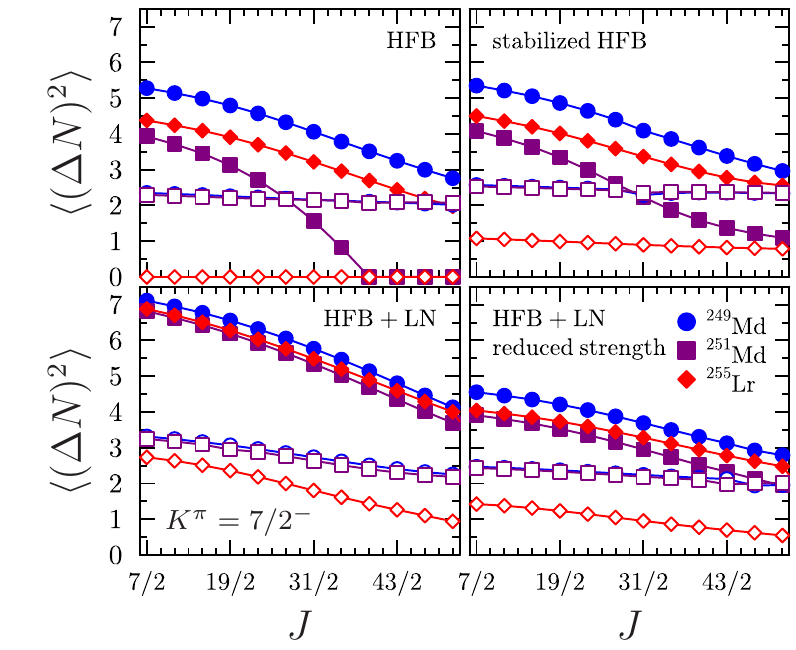} 
		}
	\end{center}
	\caption{Dispersion of neutron number (filled symbols) and proton number (open symbols)
	of states 
	obtained when blocking the favored orbit for the
	$K^\pi =  7/2^-$ band, calculated with SLy5s1 and the pairing options 
	for the three nuclei as indicated. }
	\label{fig:pair:35-}
\end{figure}

The resulting $E2$ transition energies in the 
two $K^\pi = 7/2^-$ bands
are displayed in Fig.~\ref{fig:egamma:35-}.
It is immediately visible that the calculated energies depend significantly on the pairing option.
 To understand the origin of the differences between 
pairing options and nuclei, 
Fig.~\ref{fig:pair:35-} displays
the corresponding dispersion of particle 
number $\langle (\Delta N)^2 \rangle = \langle N^2 \rangle -  \langle N \rangle^2$. The latter is  a
measure for the amount of pairing correlations.

Within a given pairing scheme, all calculated bands are very similar at low spin. There are, 
however, visible differences between the actual transition energies when comparing the four 
pairing schemes. For these nuclei that all are in the weak-pairing limit for either protons or 
neutrons or both, using stabilized HFB or HFB+LN instead of pure HFB reduces the moment 
of inertia when the calculations were done with the same pairing strength, as these schemes tend
to enhance pairing correlations. For the lowest transitions, the best agreement between the bands in
different nuclei is found for HFB, but at higher $J$ the bands visibly differ for that scheme, 
in particular the one of $^{251}$Md. This is a consequence 
of the breakdown of neutron pairing with increasing spin, which quickly increases the moment of 
inertia for this nucleus. Preventing the collapse of pairing with any of the other three pairing schemes 
brings the transition energies much closer together over the entire band. It is to be 
noted that the breakdown of neutron pairing at high spin in $^{251}$Md is an artifact of the
too large $N=150$ gap at the Fermi energy visible in Fig.~\ref{fig:nilsson:n}. Similarly, the
breakdown of proton pairing in the HFB calculation of $^{255}$Lr is an artifact of the too large
$Z=104$ gap. Assuming that the deformed gaps were at $N=152$ and $Z=100$ instead, the 
relative amount of pairing correlations would be quite different: Protons should be more paired in
$^{255}$Lr than in the two Md isotopes, while neutrons should be less paired in $^{255}$Lr than
the Md isotopes.

\begin{figure}[htb]
	\begin{center}
		\resizebox{0.48 \textwidth}{!}{
			\includegraphics{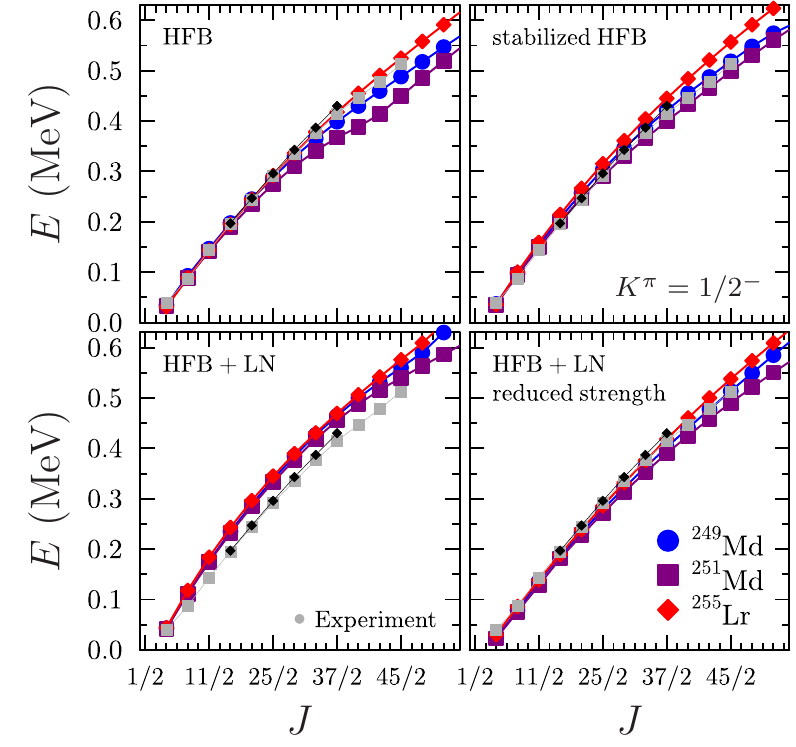} 
		}
	\end{center}
	\caption{Same as Fig.~\ref{fig:egamma:35-}, but 
	blocking the favored orbit for the
	$K^\pi = 1/2^-$ band.}
	\label{fig:egamma:05-}
\end{figure}

\begin{figure}[htb]
	\begin{center}
		\resizebox{0.48 \textwidth}{!}{
			\includegraphics{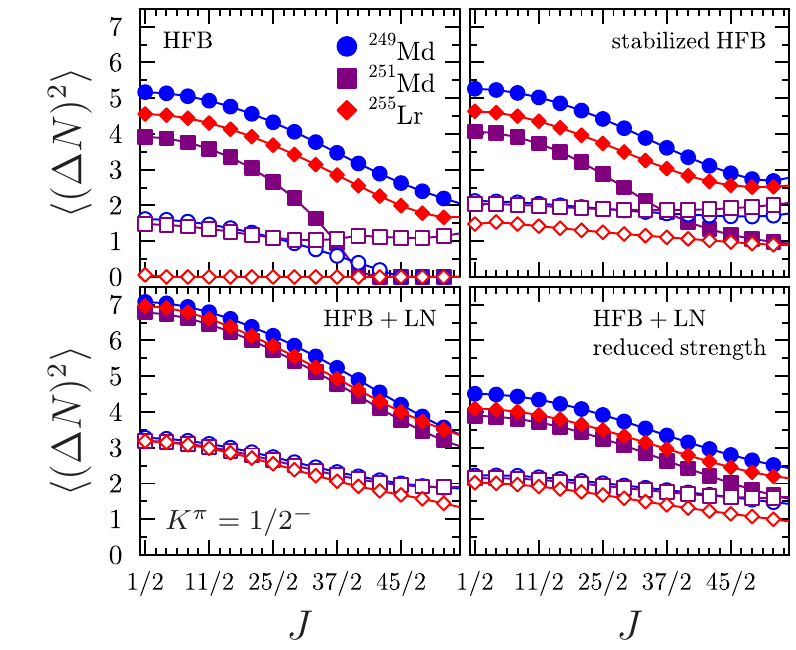} 
		}
	\end{center}
	\caption{Same as Fig.~\ref{fig:pair:35-}, but 
	blocking the favored orbit for the
	$K^\pi = 1/2^-$ band.}
	\label{fig:pair:05-}
\end{figure}

Figure~\ref{fig:egamma:05-} displays the transition energies between levels in
the $K^\pi=1/2^-$ band of the same three nuclei, and 
Fig.~\ref{fig:pair:05-} displays
the corresponding 
dispersions of particle number. The overall trends
are very similar to what is found for the $K^\pi=7/2^-$ bands. Again, the very close 
agreement of transitions in HFB at low spin is spoiled when neutron pairing breaks down at 
higher spin, an effect that is visibly reduced when using stabilized HFB or the LN scheme, in
particular at high pairing strength. It is noteworthy that the similarity of
the three calculated $K^\pi=1/2^-$ bands is slightly better than the 
agreement between the 
three calculated
$K^\pi=7/2^-$ bands,
while for data this is the other way round. Differences between the
transition energies between same levels in the different $K^\pi=1/2^-$ bands are nevertheless
still larger than what is found in experiment by about a factor of 2.

In spite of the wrong relative order of the $1/2^-$ and $7/2^-$ proton levels, the down-sloping
$1/2^-$ levels are almost empty in the calculated excited $7/2^-$ band of the Md isotopes, while 
they are almost completely filled for $^{255}$Lr as one would expect if the level sequence were 
the one suggested by experiment as depicted in Fig.~\ref{fig:orbitals}. Similarly, the up-sloping $7/2^-$ levels are almost empty for 
the $1/2^-$ band of the Md
isotopes, whereas they are almost completely empty for $^{255}$Lr as would be expected from
the empirical shell structure. As proton pairing is weak for these odd-$Z$ nuclides anyway, the 
blocked proton configurations are therefore not much affected by the imperfections of the 
single-particle spectrum.

All states in the calculated rotational bands have a dimensionless quadrupole 
deformation as defined in Ref.~\cite{Rysssens2019a} of $\beta_{2} \simeq 0.3$, 
with differences on the few percent level that depend 
on the nucleus, spin, blocked state, and pairing option used. 
With increasing spin $J$, the deformation of all configurations is slowly 
decreasing. In parallel, all configurations become slightly triaxial, with 
the $\gamma$ angle remaining below 2 deg. Comparing bands, we observe some 
systematic differences in quadrupole deformation that can be attributed to
differences in the filling of single-particle levels near the Fermi energy.
The $\beta_{2}$ value of the $7/2^-$ band of the two Md isotopes is slightly 
smaller by about 0.003 than the $\beta_{2}$ value of the $7/2^-$ band of 
$^{255}$Lr for all pairing options but HFB+LN. This is a consequence of the 
two additionally filled down-sloping, and therefore deformation-driving, $1/2^-$
levels as already discussed for the schematic Nilsson diagram of 
Fig.~\ref{fig:orbitals}. The enhanced proton pairing correlations produced by the 
HFB+LN scheme reduce this effect and lead to a near-identical deformation of 
the $7/2^-$ band for all three nuclei.
Similarly, the calculated deformation of the $1/2^-$ band of the Md isotopes 
is systematically larger than the deformation of the $7/2^-$ band. The difference 
$\Delta \beta_{2}$ is as large as 0.006 for the HFB option but remains much 
smaller for the standard HFB+LN scheme. This can be attributed to the filling 
of the deformation-driving $1/2^-$ level, while the up-sloping $7/2^-$ is 
almost empty. The deformation of the $1/2^-$ bands of the Md isotopes
is also larger than the deformation of the $1/2^-$ band of $^{255}$Lr because 
the filled up-sloping $7/2^-$ levels in the latter drive the shape to smaller deformations. The effect is again largest with a $\Delta \beta_{2}$ of about 
0.012 when using the HFB option that does not produce proton pairing correlation 
for $^{255}$Lr such that the change in the filling of orbits is largest. Using the
other pairing schemes, the $7/2^-$ level is always partially filled to a varying
degree, such that the change in deformation is reduced to about half that size.

The self-consistent calculations thereby confirm the schematic analysis of 
Fig.~\ref{fig:orbitals} concerning deformation changes, including the finding that deformation
cannot be the sole explanation for the experimentally found reduction of the moment of inertia 
of both bands when going from $^{251}$Md to $^{255}$Lr, as it only brings a change into the 
right direction for the $1/2^-$ bands. Changes in pairing correlations also have to be an 
important factor. First of all, with increasing pairing correlations this simple picture of deformation 
changes driven by proton levels being filled or empty becomes blurred. Second, a reduction
of pairing correlations in general reduces in-band transition energies \cite{Rowe1970,deVoigt1983}.
As shown in Figs~\ref{fig:pair:35-} and ~\ref{fig:pair:05-}, the calculated pairing correlations are lower in $^{255}$Lr compared to  $^{251}$Md, 
which should lead to an increase (decrease) of the moment of inertia (transition energies) in $^{255}$Lr while the opposite trend is needed to reconcile the contradictions mentioned above.

To summarize the discussion, the similarity of calculated transition energies in spite of sizable differences in the other 
properties discussed above points to accidental cancellation effects between the 
changes in shell structure, deformation, and pairing as ingredients of the identical 
$K^\pi=7/2^-$ bands and near-identical $K^\pi=1/2^-$ bands in $^{251}$Md and $^{255}$Lr.
However, it is difficult to quantify the changes brought by these effects, 
such that an additional mechanism might be at play that leads to a 
universal reduction of the moment of inertia of $^{255}$Lr compared
to $^{251}$Md. Even if such a yet unidentified mechanism is needed, it is
qualitatively described by the cranked HFB calculations, at least at low spin.
With
increasing spin, the differences between the calculated bands become larger, as is the case
for experiment.
The calculations predict that the respective band of $^{249}$Md will also be very similar to 
what was found for $^{251}$Md and $^{255}$Lr, again in spite the large differences between 
deformation and pairing. The sensitivity of the calculated transition energies to details 
of the pairing scheme also suggests that obtaining identical bands to a precision that is comparable with experiment is essentially a fine-tuning problem.
Using the SLy4
parametrization instead of SLy5s1 produces slightly different
results but leads to the same conclusions.

All of these conclusions have to remain qualitative, though, as it should not be forgotten that finding identical bands at the 1-keV level is beyond the limits of what can be expected for the systematic 
errors of the cranked HFB method as such. It is also difficult to assess the possible role of
octupole correlations, whose presence is hinted by the present data as
	discussed in Sec.~\ref{sec:octupole}, on
the values for transition energies, as the coupling of states with octupole phonons
is outside of the scope of any pure mean-field model.
As a first step in that direction, exploratory beyond-mean field calculations including particle-number
and angular-momentum projections on top of (parity-conserved) triaxial
one-quasiparticle states were recently performed for $^{251}$Md,  
using a variant of the Skyrme EDF designed for this
particular purpose~\cite{HeenenPaul-Henri2016}.
Although these calculations yield moments of inertia that are too small, they appropriately predict a $K^\pi = 7/2^-$
ground state as well as the correct ordering of the levels in the
signature partner bands.

\section{Summary and conclusion}

To summarize, this work provides the detailed properties of two rotational bands in the odd-$Z$ $^{251}$Md
interpreted as built on the $[514]7/2^-$ and $[521]1/2^-$ Nilsson orbitals, the former being the g.s.~band.
Conversion electron spectroscopy allowed the rotational bands to be extended to lower rotational frequencies for the band based on the $[521]1/2^-$ Nilsson orbital.
The conversion electron intensity was also used to constrain the single-particle configuration for the $K=7/2$ band, hence excluding the $[633]7/2^+$ configuration.
It was also shown that the band intensity profile in the presence of large internal conversion oscillates,
providing a method to deduce the gyromagnetic factor.
The most intense transition in both $^{249,251}$Md has been tentatively interpreted as a $3/2^- \rightarrow 7/2^+$ $M2$ transition, its rate being probably enhanced by octupole correlations.
The observation of several identical transitions in the $^{251}$Md-$^{255}$Lr pair is the only
case identified so far for odd-mass transuranium nuclei, which moreover differ by four mass units.
Arguments based on a mass-deformation-pairing compensation fail to explain the experimental similarities ($7/2^-$) and differences ($1/2^-$) between $^{251}$Md and $^{255}$Lr.
An additional and unexplained mechanism reducing the moment of inertia in $^{255}$Lr,
that is probably independent of the filling of specific level, would explain simultaneously IBs for the $7/2^-$ configuration and even larger changes of the moment of inertia for the  $1/2^-$ bands.
HFB calculations suggest there is not a simple
mechanism leading to identical bands in the $A=250$ mass region.
Therefore, the similarity can be hence considered as accidental.
While the collective properties are generally well reproduced by the present calculations, our study of the particular case of similar
bands points to the high sensitivity 
of the model to its ingredients and in particular to pairing correlations. 
From both an experimental and theoretical point of view, 
the present work
provides a step toward a better description of the super-heavy
nuclei region and the still speculative island of stability.

\section*{Acknowledgements}

We acknowledge the accelerator staff at the University of Jyv\"askyl\"a for delivering a high-quality beam during the experiments.
Support has been provided by
the
EU 7th Framework Programme Integrating Activities - Transnational Access Project No.
262010 (ENSAR), by the Academy of Finland under the Finnish Centre of Excellence
Programme (Nuclear and Accelerator Based Physics Programme at JYFL, Contract No. 213503), 
and by the UK STFC.
We thank the European Gamma-Ray Spectroscopy pool  (Gammapool) for the loan of the germanium detectors used in the SAGE array. 
B.B. acknowledges the support of the Espace de Structure et de r{\'e}actions
Nucl{\'e}aire Th{\'e}orique (ESNT) at CEA in France. The self-consistent mean-field computations 
were performed using HPC resources of the computing center of the IN2P3/CNRS.
W.R. gratefully acknowledges support by U.S. DOE grant No.
DE-SC0019521

\hypersetup{
	colorlinks=true,
	linkcolor=blue,
	urlcolor=blue,
	citecolor=blue
}
\bibliography{md251_2019}

\end{document}